\documentclass[floatfix,twocolumn,aps,prb,showpacs,superscriptaddress]{revtex4}
\usepackage{amssymb}
\usepackage{graphicx}
\usepackage{dcolumn}
\usepackage{bm}

\begin{document}

\title{Thermal transport in antiferromagnetic spin-chain materials}
\author{A. L. Chernyshev}
\affiliation{Department of Physics, University of California, Irvine,
  California 92697}
\author{A. V. Rozhkov}
\affiliation{Institute for Theoretical and Applied Electrodynamics
JIHT RAS, Moscow, ul. Izhorskaya 13/19, 127412, Russian
Federation}
\date{\today }

\begin{abstract}
We study the problem of heat transport in one-dimensional (1D)
spin-chain systems weakly coupled to three-dimensional phonons and
impurities. We consider the limit of fast spin excitations and
slow phonons, applicable to a number of compounds of the cuprates
family, such as Sr$_2$CuO$_3$, where the superexchange $J$ is much
larger than the Debye energy, $\Theta_D$. In this case the Umklapp
scattering among the spin excitations is strongly suppressed for
all relevant temperatures. We argue that the leading scattering
mechanism for the spin excitations at not too low temperatures is
the ``normal'' (as opposed to the Umklapp) spin-phonon scattering
in which the non-equilibrium momentum is transferred from the spin
subsystem to phonons where it quickly relaxes through the ``phonon
bath''. Because of the lower dimensionality of the spin
excitations it is only the momentum along the chains which is
conserved in such a scattering. We find that this effect leads to
a particular momentum- and temperature-dependence of the
spin-phonon relaxation rate valid for the broad class of
low-dimensional spin systems. Subsequently we demonstrate that the
spin-phonon relaxation mechanism is insufficient for the
low-energy, long-wavelength 1D spin-chain excitations, which make
the thermal conductivity diverge. We complete our consideration by
taking into account the impurity scattering, which in 1D cuts off
the quasi-ballistic
spin excitations and renders the thermal conductivity finite.
Altogether, these effects yield the following spin-boson
thermal conductivity  behavior:
$\kappa_s\propto T^2$ at low temperatures, $\kappa_s\propto T^{-1}$
at intermediate temperatures, and $\kappa_s=const$ at
higher temperatures $T\sim\Theta_D$.
The saturation at higher temperatures is of rather non-trivial origin
and we provide a detailed discussion of it. Our results compare very well
with the existing experimental data for Sr$_2$CuO$_3$. Using our microscopic
insight into the problem we propose further experiments and predict an
unusual impurity concentration dependence for a number of quantities.
\end{abstract}

\maketitle
\section{Introduction}
{\it Experiments.}\ \
Recent experiments in low-dimensional quantum magnets
\cite{Hess,Hess1,Hess1a,Sologubenko,Sologubenko1,Sologubenko_1a,Sologubenko2,Sales,Sales1,Sales2,Kudo,Takeya,Hofmann,Hess2,Nakamura,Hucker,Ando7,Hofmann1}
 have revealed remarkably strong thermal transport anomalies
associated with the low-dimensional spin degrees of freedom.
In particular, an anisotropic thermal conductivity, comparable in
magnitude to that of metallic systems,
was observed in quasi-one-dimensional (1D)
spin-$\frac{1}{2}$ chain and ladder compounds
with a large part of the heat current attributed to magnetic
excitations.\cite{Hess,Hess1,Hess1a,Sologubenko,Sologubenko1,Sologubenko_1a,Sologubenko2}
In the quasi-2D and some quasi-1D materials various other effects
were observed including strong increase of the
thermal conductivity by a modest magnetic
field\cite{Sales,Sales1,Sales2,Kudo} and enhanced
scattering due to suppression of
the gap in the spin-gaped materials.\cite{Takeya,Hofmann}
The thermal transport anomalies in layered cuprates associated with
the magnetic and/or stripe excitations have also been
found.\cite{Hess2,Nakamura,Hucker,Ando7,Hofmann1}

Both the strength and the temperature range for these new
effects are very different from their counterparts in 3D magnetic
materials, where the heat transport by well-defined magnetic excitations
exist only well below the corresponding 3D ordering
transition temperature.\cite{Sato,Douglass,Sanders}
Note that the thermal transport in low-dimensional magnets
has been studied before\cite{Lang,Buys,Gronckel} and the
qualitative reasons for that difference has been understood.
In 2D systems magnetic excitations can be very well
defined in a paramagnetic phase
because of the large in-plane magnetic correlation
length\cite{CHN} and in 1D systems the soliton-like
excitations can exist without any long-range order.\cite{Mikeska}

{\it Theories.}\ \ However, only recently a considerable progress
in the theoretical understanding of the  transport properties of
1D quantum spin systems has been
made.\cite{Narozhny,Prelovsek,Zotos,Zotos1,Alvarez,Cabra,Alvarez1,Saito,Saito1,Orignac,Andrei,Andrei1}
The main focus of these recent studies has been on the
relationship of the  spin transport and conservation
laws,\cite{Narozhny,Prelovsek} specifically on the possibility of
an ideal conducting state in one-dimensional integrable and
non-integrable systems.\cite{Alvarez,Cabra,Alvarez1} While this
problem is of significant interest, the importance of the
spin-phonon and spin-impurity couplings, which break down the
integrability of the underlying spin-only models, has also been
discussed.\cite{Andrei,Andrei1,Orignac} In particular, it was
obtained within the memory matrix formalism that the interplay of
Umklapp scattering and spin-phonon coupling leads to the
exponential temperature dependence of the thermal conductivity
$\kappa \propto e^{T^*/T}$, with $T^*$ proportional to the
phonon's Debye energy $\Theta_D$.\cite{Andrei,Andrei1}

{\it Boltzmann equation approach.}\ \
In the present work\cite{us_PRL} we consider the problem
of anomalous heat transport in quasi-1D spin-chain systems
under a somewhat different prospective.
First, we derive microscopically a model of 1D spin-boson excitations
interacting with the 3D phonon environment and impurities, where
the spin-boson representation of the 1D Heisenberg model is obtained by
performing the standard Jordan-Wigner transformation followed by
bosonization.\cite{book} It is assumed that the spin-boson velocity is
large in comparison with the phonon velocity, $v \gg c$.
The limit $v\gg c$  ($J\gg\Theta_D$) corresponds to the
experimental situation in Sr$_2$CuO$_3$, SrCuO$_2$,
(La,Ca,Sr)$_{14}$Cu$_{24}$O$_{41}$, and other cuprate
materials\cite{Johnston,Junod} where $J\sim 2000$K and $v/c\sim
10$. This large difference in spin and phonon energy scales is
also responsible for the fact that, experimentally, the ``spin
 peak'' in thermal conductivity occurs at the temperatures
well above the ``phonon
peak''.\cite{Hess,Hess1,Hess1a,Sologubenko,Sologubenko1,Sologubenko_1a,Sologubenko2}
With this assumption and in the limit of weak spin-lattice
coupling we solve the Boltzmann equation for the spin-boson
distribution function and find the spin-phonon relaxation time
$\tau_{sp}$.

{\it Normal vs Umklapp.}\ \ Why $\tau_{sp}$ is important?
Generally, the relaxation of the heat current and the resulting
finite thermal conductivity should be due to the Umklapp processes
or any other processes which do not conserve momentum. Since the
characteristic ``bandwidth'' for the spin excitations in the real
1D spin chains materials of interest is very large ($J\sim 2000$K)
in comparison with the experimental temperature range, the Umklapp
scattering of spin excitations on themselves is strongly
suppressed (as $\sim e^{-J/T}$). Instead, by considering the limit
$v/c \gg 1$, we propose that the leading relaxation mechanism at
not too low temperatures is the two-stage, bottle-neck process of
{\it (i)} transferring momentum from the spin system to phonons,
and {\it (ii)} subsequent dissipation of the phonon momentum via
an Umklapp process or impurity scattering. The central idea is
that the excess momentum of a spin-boson waits the longest time to
get transfered to phonons, but once it is transferred it relaxes
quickly. In other words, the phonon relaxation time $\tau_{pp}$
due to phonon-phonon scattering is much shorter than the
spin-phonon relaxation time $\tau_{sp}$, i.e. $\tau_{pp}\ll
\tau_{sp}$. As a result, the relaxation rate of such a two-stage
process is determined by the ``normal'' spin-phonon scattering
rate
$\tau_{2-st}^{-1}=[\tau_{sp}+\tau_{pp}]^{-1}\approx\tau_{sp}^{-1}$.
The  inequality $\tau_{pp} \ll \tau_{sp}$ can be justified
qualitatively with the help of the following argumentation. First,
the spin-lattice coupling is weak. Second, since $c\ll v$ the
phonon states have a significant thermal population at the
temperatures where the spin contribution to the transport is
substantial.

Furthermore, there is also a kinematic argument in favor of $\tau_{pp}\ll
\tau_{sp}$  that results from our analysis.
The leading contribution to the transport comes from
the spin bosons with a small momentum $k \ll T/v$.
We find that the most effective scattering mechanism for such bosons
is an absorption of a ``thermal'' phonon, that is a
phonon whose energy is of the order of $T$ and momentum is of the
order $P_T= T/c \gg T/v \gg k$. For finite $T$ such phonons have finite
relaxation time $\tau_{pp}^T$. We will show that the spin bosons in
question have a divergent relaxation time $\tau_{sp}(k) \propto 1/k^2$.
Therefore, for spin bosons with sufficiently small $k$'s
the relevant phonon relaxation time $\tau_{pp}^T$
is shorter than the spin-phonon relaxation time $\tau_{sp}$  not only
because of the weak spin-phonon coupling, but also due to the smallness
 of the characteristic momentum of the spin-boson, $P_T \gg k$.

We have performed a detailed study of the
kinematics of the spin-phonon scattering processes and obtained the
corresponding transport relaxation rate
$\tau_{sp}^{-1}\simeq {\cal A}k^2 T^3/v^3$ for $T\ll \Theta_D$,
and $\tau_{sp}^{-1}\simeq {\cal A}k^2 T \Theta_D^2/v^3$ for $T\agt \Theta_D$,
 ${\cal A}$ is a constant
related to the spin-phonon coupling, ${\cal A}\propto (g_{sp}/c)^2$.
The characteristic power of $k^2$ in the relaxation rate can be traced
back to the fact that spin excitations are confined to lower dimensions
than phonons. The latter implies that it is only the momentum along the
chains which is conserved in the spin-phonon scattering. This, in turn,
leads to a larger scattering space for phonons and makes relaxation
rate of the spin bosons
$\propto k^2$. Note that for the problem of the
phonon-phonon scattering in 3D the
phonon scattering rate is $\propto k^4$ because the
scattering space is restricted by the momentum conservation in all three
dimensions.\cite{Ziman}
The temperature dependence of the relaxation rate is due to the bosonic
nature of both the spin excitations and the phonons.
Thus, the spin-phonon relaxation rate obtained in this work should
apply to other low-dimensional spin systems as well.

{\it Infrared divergence.}\ \
However, one can find that
the considered scattering mechanism becomes too weak at low energy
and ineffective for the dissipation of the
low-energy spin bosons. Namely, within the formalism of Boltzmann
equation, $\tau_{sp}^{-1}\propto k^2$ leads to the infrared
diverging thermal conductivity $\kappa_s\propto\int dk/k^2\sim
k^{-1}|_{k_0}$, $k_0\rightarrow 0$,
the situation familiar from the phonon thermal
conductivity in 3D insulators.\cite{Ziman,Mahan,Berman,Bhandari}
In that latter 3D phonon problem the regularization of such a divergence is
non-trivial and involves a consideration of the
higher-order phonon processes,\cite{Mahan,Chuk} degeneracy of phonon
branches at high
symmetry points,\cite{Herring} or scattering  on boundaries.

{\it Impurities.}\ \
What physical effect can render $\kappa_s$ finite in our case?
In contrast with the 3D systems, \cite{erdos} because of the 1D nature
of spins, even a weak impurity potential will have a
dramatic effect on the low-energy spin bosons.
The impurities generate a relevant interaction in the renormalization
group (RG) sense. That is, they scatter low-energy excitations very
well destroying their quasi-ballistic propagation. Below the so-called
Kane-Fisher scale $T_{KF}$ the scattering is very strong
and leads to the localization of 1D excitations.\cite{KaneFisher}
On the other hand, at the temperatures well above this scale
$T\gg T_{KF}$ the impurity scattering can be analyzed perturbatively. As a
consequence of such an analysis we find that impurities
result in the momentum-independent scattering rate
$\tau_{imp}^{-1}\propto n T^{-1}$, $n$ being the impurity concentration.
Total scattering rate $\tau^{-1}_{tot} = \tau^{-1}_{sp} + \tau^{-1}_{imp}$ is
finite at $k=0$, which makes $\kappa_s$ finite.

{\it Results.}\ \
Altogether, the combined effect of impurities and the spin-phonon
scattering leads to the following spin-boson thermal conductivity
behavior: {\it (i)} in the low-temperature, impurity-scattering
dominated regime $\kappa_s~\propto~T^2$,
{\it (ii)}  in the spin-phonon scattering dominated
regime, intermediate temperatures $T_m \ll T\ll\Theta_D$,
$\kappa_s\propto  T^{-1}$, and {\it (iii)}
 in the spin-phonon scattering dominated regime, higher temperatures
 $T\agt\Theta_D$, $\kappa_s~=~const~\propto~T_m/\Theta_D$.
Here the temperature $T_m$ corresponds to the maximum in $\kappa_s(T)$.
It is also the  crossover temperature between the impurity- and
phonon-scattering dominated regimes.

We find that this temperature behavior of the thermal conductivity
agrees very well with the available experimental data for the
spin-chain material
Sr$_2$CuO$_3$.\cite{Sologubenko1,Sologubenko_1a} The thermal
conductivity for the zig-zag chain SrCuO$_2$ and the spin-ladder
compounds
(Ca,La,Sr)$_{14}$Cu$_{24}$O$_{41}$,\cite{Sologubenko1,Hess} whose
underlying spin models are different from the ones considered in
this work, is briefly discussed.

We analyze the impurity concentration dependence of several
quantities. As a result we predict an unusual behavior of:
{\it (i)} the crossover temperature $T_m\propto n^{1/6}$,
{\it (ii)} the spin-boson thermal conductivity maximum value
$\kappa_s^{max}=\kappa_s(T_m)\propto n^{-2/3}$,
and {\it (iii)} the asymptotic value of $\kappa_s$
at $T \gg T_m$: $\kappa_s^{\infty}\propto n^{-1/2}$.

{\it Outline.}\ \
Our paper is organized as follows.
In Sec. \ref{Hamiltonian}
we introduce the spin-chain Hamiltonian and derive the spin-phonon and
spin-impurity interaction terms.
In Sec. \ref{phonons} we find the spin-boson
relaxation time due to scattering off the phonons. In
Sec. \ref{impurities}
the impurity contribution to the spin-boson relaxation time is calculated.
In Sec. \ref{kappa} the thermal conductivity {\it vs} temperature
is obtained and the results are compared to experimental data. In
Section \ref{kappa} we
also put forward several theoretical predictions for
the impurity concentration dependence of different quantities and
suggest further experiments. In Sec. \ref{meanfp} we discuss the notion
of the mean free path and its applicability to the transport by the
long-wavelength spin bosons.
We conclude by Sec. \ref{Conclusions} which contains the discussion of our
approach and approximations.
Technically involved details concerning the spin-phonon collision
integral and impurity scattering are described in Appendix \ref{app_A}
and Appendix \ref{app_B}, respectively.

\section{Spin-phonon and spin-impurity interaction Hamiltonians}
\label{Hamiltonian}

{\it Spin-chain Hamiltonian.} \ \
The Hamiltonian of a single Heisenberg antiferromagnetic spin-$\frac12$
chain is:
\begin{eqnarray}
H_{\rm chain} = J \sum_i {\bf S}_i \cdot {\bf S}_{i+1}.
\end{eqnarray}
This Hamiltonian, after the Jordan-Wigner transformation, can be expressed
in terms of fermionic operators $\psi_i$ as follows:
\begin{eqnarray}
H_{JW} &=& -\frac{J}{2} \sum_i \bigg(\psi^\dagger_i
\psi^{\vphantom{\dagger}} _{i+1} + {\rm h.c.} \\
&& +\frac12
\left(2\psi^\dagger_i \psi^{\vphantom{\dagger}} _i -1\right)
\left(2\psi^\dagger_{i+1}
\psi^{\vphantom{\dagger}} _{i+1} - 1\right)\bigg) \ .\nonumber
\end{eqnarray}
It is convenient to bosonize this Hamiltonian. To do that one introduces
chiral fermionic fields $\psi_{L,R}(x)$ such that $\psi (x) \approx
\psi_{L}(x) e^{ik_F x} + \psi_{R}(x) e^{-ik_F x}$. These fields can be
rewritten in terms of spin boson field $\Phi(x)$ and its dual field
$\Theta(x)$ as
\begin{eqnarray}
\psi_{L,R} = \frac{1}{\sqrt{\pi a}} e^{i\sqrt{\pi}(\Theta \pm \Phi)}.
\end{eqnarray}
Using $\Phi, \Theta$ one writes the spin-chain Hamiltonian as:
\begin{eqnarray}
\label{H00}
H_0 = \frac{v}{2}
\int dx \left({\cal K}
\left( \partial_x \Theta \right)^2 + {\cal K}^{-1}
\left( \partial_x \Phi \right)^2 \right),
\end{eqnarray}
where $\int dx\equiv \int_{-L/2}^{L/2} dx$, $L$ is the linear size of
the system, the Luttinger-liquid  parameter ${\cal K}$ is equal to
$1/2$, and  the spin-boson velocity $v$ is given by
\begin{equation}
v =\frac{\pi}{2} Ja,
\end{equation}
where $a$ is the chain lattice constant. Furthermore, the
spin-boson variables are redefined according to the rules:
\begin{eqnarray}
&&\tilde \Theta = {\cal K}^{1/2}\Theta,\\
&&\tilde \Phi = {\cal K}^{-1/2} \Phi.\nonumber
\end{eqnarray}
Then we introduce the creation and annihilation operators for
the field $\tilde \Phi(x)$:
\begin{eqnarray}
\label{Phi}
\tilde \Phi (x) = \sum_k \frac{e^{ikx}}{\sqrt{2L|k|}} \left( b^\dagger_k
+ b^{\vphantom{\dagger}}_{-k} \right)\ ,
\end{eqnarray}
which diagonalize the Hamiltonian $H_0$, Eq. (\ref{H00}):
\begin{eqnarray}
H_0 = {v} \sum_k |k| b^\dagger_k b^{\vphantom{\dagger}}_k.\label{H0}
\end{eqnarray}
For details of this procedure see Ref. \onlinecite{book}. 
We would like to note that the Hamiltonian for the spin-$\frac12$
chain contains other terms beyond the LL form above. These terms do
not lead to the finite heat conductivity by themselves 
and thus are not included in our consideration. 
The role of such terms in the presence of phonons
via a phonon phonon-assisted Umklapp process has been considered 
elsewhere.\cite{Andrei}

{\it Spin-Phonon Hamiltonian.} \ \
Since the superexchange $J$ is a function of inter-site separation,
the lattice vibrations are able to modify it. This mechanism
will, therefore, couple phonons to the spin degrees of freedom.
We can account for this coupling by expanding  $J$ in
gradients of inter-site distance $a\partial_x u_x$:
\begin{equation}
J (a + a\partial_x u_x) = J(a) + \delta J \partial_x u_x + \ldots
\end{equation}
where $u_x$ is the displacement along the chain.
This translates into the following interaction Hamiltonian:
\begin{eqnarray}
\label{Hsp}
H_{\rm int} = \frac{ g_{sp}}{2} \int dx \left({\cal K}
\left( \partial_x \Theta \right)^2 + {\cal K}^{-1}
\left( \partial_x \Phi \right)^2 \right)\left( \partial_x u_x \right),
\end{eqnarray}
where $g_{sp}=\delta Ja$ is a spin-phonon coupling constant.
The atomic displacement vector ${\bf u}$ can be
expressed in terms of phonon creation and annihilation operators as follows:
\begin{eqnarray}
\label{u}
{\bf u}({\bf R}) = \frac{1}{\sqrt{N}}
\sum_{{\bf P} \ell} \frac{e^{i{\bf PR}}}{\sqrt{2m_i
\omega_{{\bf P}\ell}}} \ \
{\bm \xi}_{{\bf P}\ell} \left(a^\dagger_{{\bf P}\ell}
+a^{\vphantom{\dagger}}_{{\bf -P}\ell} \right) \ ,
\end{eqnarray}
where $\omega_{{\bf P}\ell} = c_\ell |{\bf P}|$ and
summation runs over the three-dimensional wave-vector ${\bf P}$ and
over three polarizations of 3D phonons denoted by $\ell$. Here
${\bm \xi}$ is the polarization vector of a phonon, $m_i$ is the
mass of the unit cell,
$N$ is the number of unit cells in the sample, ${\bf R} = (x, y,
z)$ is the position vector. Sound velocities $c_\ell$ are much smaller
than the spin-boson velocity $v$:
\begin{eqnarray}
c_\ell \ll v \ . \label{c<v}
\end{eqnarray}
For a chain specified by $y=0, z=0$ the interaction Hamiltonian,
Eq. (\ref{Hsp}), can be written, using Eqs. (\ref{u}) and
(\ref{Phi}), as:
\begin{eqnarray}
H_{\rm sp} = -\frac{i g_{sp}}{\sqrt{N}} \sum_{kk'{\bf P}\ell}
V_{\ell}({\bf P},k,k')
\left( a^\dagger_{{\bf P}\ell} b^\dagger_k b^{\vphantom{\dagger}}_{k'}
 + {\rm h.c.} \right)\ ,\label{Hint}
\end{eqnarray}
with
\begin{eqnarray}
V_{\ell}({\bf P},k,k')=
\frac{P_\| k k'}{\sqrt{ 8m_i \omega_{{\bf P}\ell}kk'}}
\left({\bm\xi}_{{\bf P}\ell}\right)_x\delta_{P_\|,k'-k}\ ,\label{Vint}
\end{eqnarray}
where $k$ and $k'$ are the 1D momenta of spin bosons and ${\bf P}$ is the
3D momentum of a phonon. Note that only component of the
total momentum along the chain is conserved, which is explicitly given by
the $\delta$-symbol.
The projection of the polarization vector on the $x$-axis
$( {\bm\xi}_{{\bf P}\ell})_x$ is equal to:
\begin{eqnarray}
\label{xi}
\left( {\bm \xi}_{{\bf P}\ell}\right)_x = \cases{
|P_\| |/|{\bf P}|& longitudinal,\cr
\sqrt{1-(P_\|/{\bf P})^2}& transverse,\cr
0 & transverse,\cr}
\end{eqnarray}
where two answers for the transverse phonons correspond to two possible
choices of polarization ${\bm\xi}$. It is convenient to choose the first
of these polarizations to lie in the plane given by $\hat x$ and ${\bf P}$
and the second one to be normal to this plane. Since the projection
${\bm\xi}_x$ is zero in the second case, such phonons do not couple to the
spin bosons  and are not discussed in this paper any further.

Altogether,
the Hamiltonian (\ref{Hint}) describes emission and absorption of
phonons by the spin-density excitations represented by the spin bosons, see
Fig. \ref{phonon_diagrams}.

{\it Spin-Impurity Hamiltonian.} \ \
In addition to interacting with phonons  spin bosons are also scattered on
impurities. Microscopically, the origin of that effect is simple:
impurity leads to a local variation of the superexchange coupling
$J$, which leads to scattering of magnetic excitations.
The impurities affect the low temperature transport properties of the
Tomonaga-Luttinger liquid in a dramatic way
because they  act as almost ideal backward scatterers for
the low-lying excitations.
The most important (relevant in RG sense) part of the impurity
Hamiltonian for an impurity located at $x_{0}$ can be written as:
\begin{eqnarray}
\label{Himp}
H_{{\rm imp} }& =&a \delta J_{{\rm imp} }
e^{-i k_{\rm F} x_0 } \psi_L^\dagger (x_0)
\psi_R^{\vphantom{\dagger}} (x_0) + {\rm h.c.} \\
&=&\frac{\delta J_{{\rm imp} }}{\pi }
\cos \left( 2k_{\rm F} x_0 + \sqrt{2\pi} \tilde\Phi(x_0 )
\right).\nonumber
\end{eqnarray}
Although the effect of impurities is very strong for the low-lying
excitations, for small $\delta J_{imp} \ll J$ and not too low temperatures
the impurity scattering can be considered
perturbatively.

\section{Phonon mechanism of relaxation}
\label{phonons}

{\it Boltzmann equation.}\ \
In order to study the relaxation of spin excitations on phonons
we will solve the Boltzmann equation for spin bosons
coupled to the bath of 3D phonons.
The stationary Boltzmann equation has the form:
\begin{eqnarray}
v \partial_x f_k = -S_k[f], \label{BE}
\end{eqnarray}
where $f_k$ is the spin-boson distribution function and $S_k[f]$ is the
collision integral. In general, the collision integral is a non-linear
functional of both $f_k$ and the phonon distribution function
$n_{{\bf P}\ell}$: $S_k = S_k [f,n]$. However, as we argued in the
Introduction, one can assume that the
relaxation of phonons is quicker than that of the spin bosons.
Thus, even in the presence of the
temperature gradient $\partial_x T$, phonons will be treated as if they
are in a local equilibrium with themselves and this equilibrium is
characterized by a local temperature $T(x) = T_0 + x\partial_x T$:
\begin{eqnarray}
n_{{\bf P}\ell} = n_{{\bf P}\ell}^0\left(T(x)\right) =
\frac{1}{e^{\omega_{{\bf P}\ell}/T(x)}-1}.
\end{eqnarray}
Therefore, one can write $S_k \approx S_k [f,n^0] = S_k [f]$.

{\it Collision integral.}\ \
The collision integral for spin bosons can be generally written as:
\begin{eqnarray}
\label{CE}
S_k= \int_{k'}
\bigg[ W_{kk'}f_k (f_{k'} + 1)- W_{k'k}f_{k'}(f_k +
  1)\bigg]\ ,
\end{eqnarray}
where $\int_{k'}$ stands for $\int dk'/2\pi$ and
$W_{kk'}$ is the total probability of the spin excitation
to be scattered from the state $k$ to the state $k'$.
For the processes of scattering due to
phonons such probabilities are given by:
\begin{eqnarray}
\label{Wkk}
W_{kk'}=\sum_\ell \int_{{\bf P}}\bigg( w^\ell_{kk'{\bf P}}
(n^0_{{\bf P}\ell} + 1) + w^\ell_{k'k{\bf P}} n^0_{{\bf P}\ell}\bigg)\ ,
\end{eqnarray}
where  the symbol $\int_{{\bf P}}$ stands for $(1/2\pi)^3\int  d^3 {\bf P}$.
Similar expression for the probability $W_{k'k}$
of scattering from $k'$ to $k$
can be obtained by permutation of $k$ and
$k'$ in (\ref{Wkk}).
Here the ``elementary'' scattering probabilities $w^\ell_{kk'{\bf P}}$
of the spin boson due to emission or absorption of the
phonon with the momentum ${\bf P}$ and polarization $\ell$
are determined from Eqs. (\ref{Hint}), (\ref{Vint}):
\begin{eqnarray}
w^\ell_{kk'{\bf P}}& =& \frac{g_{sp}^2V_0}{8m_i}\cdot
\frac{P_\|^2|kk'|}{\omega_{{\bf P}\ell}} ({\bm \xi}_\ell)^2_x
\label{w}\\
&&\times \delta(k'+P_\|-k) \delta (\omega_{k'} + \omega_{{\bf P}l}-\omega_k),
\nonumber
\end{eqnarray}
where $V_0$ is elementary cells volume, $\omega_k = v|k|$ is the spin-boson
energy.
In all the collisions the total energy and the projection of the
momentum along the chain is preserved. These conservation laws are
enforced by the $\delta$-functions in Eq. (\ref{w}).

{\it Linearized Boltzmann equation.}\ \
Boltzmann equation (\ref{BE}) should be solved in the presence of a
non-zero temperature gradient $ \partial_x T \ne 0$.
Assuming that the gradient $ \partial_x T$ is small one can introduce a
usual ansatz for the distribution function $f_k$ to
linearize the Boltzmann equation:
\begin{eqnarray}
&&f_k = f_{k}^0 + f_{k}^1,\label{f}\\
&&f_{k}^0\left(T\right) = \frac{1}{e^{\omega_k/T} - 1}, \nonumber
\end{eqnarray}
where $f_{k}^1$ is a non-equilibrium correction to the
equilibrium distribution function $f_{k}^0$. Function $f_k^1$ is
considered to be small.
Since the collision integral vanishes identically in an equilibrium
state when $f_k=f_k^0$ and $n_{\bf P} =n_{\bf P}^0$, one expects that
$f_{k}^1$ should be proportional to the temperature gradient
$\partial_x T$.

\begin{figure}[t]
\includegraphics[width=8cm]{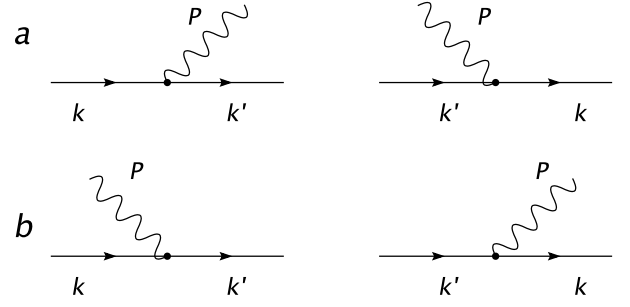}
\caption{The scattering diagrams of the spin boson on the phonon for
the processes contributing to the collision integral
(a) $S^{(1)}$ and (b) $S^{(2)}$. Solid lines are spin bosons, wavy lines are
phonons.\cite{jaxodraw}}
\label{phonon_diagrams}
\end{figure}

Thus, the Boltzmann equation for the spin bosons to the first order in
$\partial_x T$ can be rewritten as:
\begin{eqnarray}
\frac{v|k|}{T} \left(\partial_x T \right) \frac{\partial f_k^0}{\partial k}
= S_k, \label{Boltzmann}
\end{eqnarray}
where $S_k$ now stands for the linearized collision integral which can
be expressed using the notations of Eq. (\ref{f}) as:
\begin{eqnarray}
\label{Coll_int}
S_{k} &=&\sum_\ell\left(S^{(1)}_{k\ell}+S^{(2)}_{k\ell} \right)\ ,\\
\label{S1}
S^{(1)}_{k\ell} &=&
\int_{k'{\bf P}} w^\ell_{kk'{\bf P}} \bigg[ \left( n_{{\bf P}\ell}^0
+ f_{k'}^0 + 1\right) f_{k}^1  \\
&&\phantom{\int_{k'{\bf P}} w^\ell_{kk'{\bf P}}\bigg[}
+ \left(f_{k}^0-n_{{\bf P}\ell}^0 \right) f_{k'}^1 \bigg] \ ,\nonumber\\
\label{S2}
S^{(2)}_{k\ell}
&=& -\int_{k'{\bf P}} w^\ell_{k'k{\bf P}}
\bigg[ \left( n_{{\bf P}\ell}^0
+ f_{k}^0 + 1\right) f_{k'}^1  \\
&&\phantom{-\int_{k'{\bf P}} w_{k'k\ell} \bigg[}
+ \left(f_{k'}^0 -n_{{\bf P}\ell}^0 \right) f_{k}^1 \bigg]
\nonumber \ .
\end{eqnarray}
The integral $S^{(1)}$ accounts for two types of collision events:
{\it (i)} spin boson with the momentum $k$ emits a phonon with the momentum
${\bf P}$ and scatters into the state with the momentum $k'$; {\it (ii)}
process inverse to
{\it (i)}. Likewise, $S^{(2)}$ describes  the absorption of a phonon by spin
boson with the momentum $k$ and the corresponding inverse process,
see Fig. \ref{phonon_diagrams}.

Once the Boltzmann equation (\ref{Boltzmann}) is solved for $f^1_k$ the
spin-boson  thermal current density $J_E^s$ can be written as:
\begin{eqnarray}
\label{J_E}
J_E^s = \int v^2 k f^1_k \frac{dk}{2\pi} = -\kappa_s\partial_x T,
\end{eqnarray}
where the coefficient $\kappa_s$ is the spin-boson thermal
conductivity. The total thermal conductivity observed in experiment
is the sum of $\kappa_s$ and the phonon thermal conductivity.

{\it Kinematic considerations.}\ \
To solve the Boltzmann equation (\ref{Boltzmann}) one needs to evaluate
integrals in Eqs. (\ref{S1}) and (\ref{S2}). Part of
this task can be accomplished without any approximations as the
integral over ${\bf P}$ in $S^{(1)}$ and $S^{(2)}$
can be calculated explicitly.
Because of the conservation laws discussed above the remaining integration
over $k'$ will be restricted to some finite intervals whose extent and
range depend on $k$. Mathematical details of these calculations
are given in Appendix \ref{app_A}.

The collision integral derivation can be simplified even more if  we
restrict our attention to the spin bosons whose energies are small:
$v|k|\ll \min \{T; \Theta_D\}$. These energies are the most important since
the transport is dominated by the bosons with small $k$.
Such an approximation is used at various stages of $S_k$ calculation.
The consistency of our result with this latter assumption can be
verified afterward.

{\it Relaxation time.}\ \
One can write the collision integral in Eq. (\ref{Coll_int}) as a
sum of two terms:
\begin{eqnarray}
\label{2_terms}
S_{k} = \frac{f_{k}^1}{\tau_{sp}(k)}+\delta S_{k} \ ,
\end{eqnarray}
where the first term has the usual relaxation time form, while the
second one does not. However, the second term can be estimated and it
is shown to be small for small spin-boson energy $\delta S_k/S_k\propto
(vk/T)^2$ (for
details see Appendix \ref{app_A}).

Finally, with the help of these approximations, the collision integral
can be written in the relaxation time form $S_k \approx
f^1_k/\tau_{\rm sp}(k)$, where $\tau_{\rm sp}(k)$ is the transport
relaxation time. Since the transport is dominated by the spin bosons
with small momentum, one can use that smallness to simplify
significantly the corresponding expression (\ref{tau})
in the collision integral to obtain:
\begin{eqnarray}
\frac{1}{\tau_{\rm sp}(k)} = \frac{{\widetilde{\cal A}}v
  k^2}{T}\
\int_0^{\Theta_D/v} \frac{dk'}{2\pi} \  \frac{(k')^3}{2\sinh^2 (vk'/2T)}\ ,
\label{tau_tr00}
\end{eqnarray}
where ${\widetilde{\cal A}}$ is a constant proportional to the spin-lattice
coupling, $\sinh(...)$ comes from
the bosonic distribution function, and powers of $k$ and $k'$ are from
the scattering amplitudes $V({\bf P},k,k')$ in Eq. (\ref{Vint}) and due to
the integration over ${\bf P}$.

What does this formula tell us about the most effective
scattering process for a spin boson with small momentum?
One can see that the major contribution to the integral in (\ref{tau_tr00})
comes from  $vk'\sim \min \{T;\Theta_D\}$. That is, the scattered spin
boson is ``thermalized''. Therefore, the most important scattering
process is the absorption of a thermalized phonon.

Further simplification of the problem can be  achieved by exploiting
the small parameter $c/v \ll 1$. Because the spin bosons are fast, the
energy and momentum conservation dictates that the majority of the
phonons which interact with the spin subsystem must have their
momentum almost normal to the chain direction.
Indeed, the energy conservation gives that the phonon momentum
$P = \omega_{\bf P}/c = (v/c)||k|-|k'||$, where $k$ and $k'$ are the
1D spin-boson momenta, while the momentum conservation along the chain
gives $|P_\|| = |k - k'|$. Therefore, the momentum component perpendicular
to the chain for a typical phonon is: $|P_\perp| \approx P \sim
(v/c)|P_\|| \gg |P_\||$, except
for the case of almost elastic backward scattering $k' \approx
-k$. The latter
scattering event does not contribute substantially to the transport
relaxation time (see Appendix \ref{app_A}).
Since the momentum ${\bf P}$ is almost normal to
the chain the corresponding
projection of the polarization vector for the longitudinal
phonons given in Eq. (\ref{xi}) is small:
${\bm \xi}_{xl} = {\cal O} (c/v)$. At the same time the transverse
phonon polarization projection is
${\bm \xi}_{xt} = {\cal O}(1)$. Thus, the most effective spin-phonon
scattering is due to the transverse phonons.

We would like to rewrite Eq. (\ref{tau_tr00}) in the following form:
\begin{eqnarray}
\frac{1}{\tau_{\rm sp}(k)} = \frac{{{\cal A}}k^2
  T^3}{v^3}\Gamma(\Theta_D/T)\ ,
\label{tau_tr000}
\end{eqnarray}
with the auxiliary function $\Gamma(z)$ defined as:
\begin{eqnarray}
\Gamma(z)=\frac{I_1(z)}{I_1(\infty)}\ , \ \ \ \
 I_1(z)=\int_0^z \frac{x^3 dx}{2\sinh^2 x/2}\ ,
\label{gamma0}
\end{eqnarray}
and ${\cal A}$ given by:
\begin{eqnarray}
{\cal A}=\frac{I_1(\infty)}{2\pi}\ {\widetilde{\cal A}}
\propto\frac{V_0}{m_i}\left(\frac{g_{sp}}{c}\right)^2 \ .
\end{eqnarray}
$I_1(\infty)=12\zeta(3)$ where $\zeta(x)$ is the zeta-function.
One can easily verify that:
\begin{eqnarray}
\label{gamma1}
\Gamma(\Theta_D/T)=\cases{
1 & for $T\ll {\widetilde\Theta_D}$ ,\cr
({\widetilde\Theta_D}/T)^2 & for
$T\gg {\widetilde\Theta_D}$ , \cr}
\end{eqnarray}
where ${\widetilde\Theta_D}=\Theta_D/\sqrt{I_1(\infty)}\approx
\Theta_D/4$ plays the role of a crossover temperature.
That is, for the temperatures well below the Debye energy $T\ll
\Theta_D$ the integration in Eq. (\ref{tau_tr000})
is restricted by $\sinh^{-2} (x/2)$ and
the relaxation rate is given by:
\begin{eqnarray}
\frac{1}{\tau_{\rm sp}(k)} = \frac{{\cal A} T^3 k^2}{v^3} \ .
\label{tau_tr0}
\end{eqnarray}
However, for the temperatures comparable to $\Theta_D$ and higher
(in fact, for
$T\agt \Theta_D/4$) the integration limit in Eq. (\ref{gamma0}) becomes
small. This leads to the change in
the temperature dependence of the relaxation rate from $\propto T^3$
to $\propto T$:
\begin{eqnarray}
\frac{1}{\tau_{\rm sp}(k)} = \frac{{\cal A}{\widetilde\Theta_D}^2 T
 k^2}{v^3} \ .
\label{tau_tr0_a}
\end{eqnarray}
The relaxation rate for the spin boson due to scattering on  phonons
in the form given in Eq. (\ref{tau_tr000}) with the limiting cases
given by Eqs. (\ref{tau_tr0}) and (\ref{tau_tr0_a}) are the main results
of this Section.

It is instructive to inspect and compare our results with
the well-known phonon-phonon
scattering rate for the 3D phonons, studied in detail many years
ago.\cite{Ziman,Berman,Herring}
For the 3D phonon-phonon scattering rate one finds that it depends on the
higher power of the wave-vector $\propto k^4$, while its temperature dependence
is the same. This is because the scattering space in phonon-phonon case is
restricted by the momentum conservation in all three dimensions. In our case,
it is only the momentum along the chains which is conserved in the
spin-phonon scattering. This leads to fewer restrictions and a larger
scattering
space for phonons and makes the relaxation rate $\propto k^2$.
Thus, the characteristic power of $k^2$ in the relaxation rate
Eqs. (\ref{tau_tr00})-(\ref{tau_tr0_a}) can be traced
back to the fact that spin excitations are confined to lower dimensions
than phonons. This implies that our results for the spin-phonon relaxation rate
should be applicable to other low-dimensional spin systems as well.
The temperature dependence of the relaxation rate is due to the bosonic
nature of both the spin excitations and the phonons.

{\it Infrared divergence.}\ \
With the collision integral in the relaxation time form
the Boltzmann equation is trivially solved:
\begin{eqnarray}
f^1_k = \frac{v|k|\tau_{\rm sp}(k)}{T} \left(\partial_x T \right)
\frac{\partial f_k^0}{\partial k} \approx -\frac{\tau_{\rm sp}(k)}{k}
(\partial_x T)\ , \label{f1_ph}
\end{eqnarray}
where we used that $(v|k|/T) \partial f^0/\partial k \approx -1/k $
in the small $k$ limit.

Once $f^1_k$ is found the energy current density can be determined
from Eq. (\ref{J_E}) which yields the thermal conductivity:
\begin{eqnarray}
\kappa_s = \int v^2 \tau_{\rm sp}(k) \frac{dk}{2\pi}=
\frac{v^5}{2\pi {\cal A} T^3} \int_0^{T/v} \frac{dk}{k^2}\ .
\label{k_diver}
\end{eqnarray}
This expression diverges at small $k$ giving rise to an infinite $\kappa_s$.
The divergence happens because the scattering of spin
bosons on phonons is not sufficiently strong for $k \rightarrow 0$ to
ensure the convergence of the low-energy contribution to the thermal
conductivity.
Such a  situation is familiar from the phonon thermal
conductivity in 3D insulators.\cite{Ziman,Mahan,Berman,Bhandari}
In the 3D phonon problem the regularization of such a divergence is
non-trivial and involves  consideration of the
higher-order phonon scattering processes,\cite{Mahan,Chuk} degeneracy
of different phonon branches at high symmetry points,\cite{Herring} and
interface boundary scattering.

We would like to note here that such a divergence is not related to any
conservation law or integrability of the problem. The scattering
simply becomes too weak at low energies and is unable to equilibrate
the excitations. A model example of such a behavior
would be a system of free, noninteracting phonons in a continuum,
which are scattered only by the
point-like impurities according to the Rayleigh's law.\cite{erdos}
All the conservation laws are broken in this case.
Yet, the heat conductivity is infinite
in such a system since  the Rayleigh scattering is ineffective at low
energies, which leads to a diverging integral similar to our
Eq. (\ref{k_diver}).

In our case,
because of the 1D nature of spin system
the impurity scattering of spin bosons
is very effective at low energies, as we will show in the next
Section, so that even weak impurity potential renders the thermal
conductivity finite.

\section{Impurity scattering}
\label{impurities}

In a 1D system impurities have a dramatic effect on the low-energy
excitations.  The Hamiltonian for spin excitations in a 1D spin chain
interacting with a single defect is given in Eq. (\ref{Himp}). Since the
impurity Hamiltonian is a relevant perturbation in the RG sense, defects
scatters the low-energy excitations very effectively destroying their
quasi-ballistic propagation.  Therefore, the disorder in the magnetic coupling
$J$ will remove the infrared divergence in Eq. (\ref{k_diver}).

The impurity Hamiltonian in Eq. (\ref{Himp}) is not a low-order polynomial
of the bosonic field. Thus, it is impossible to account for the impurity
scattering within the formalism of the Boltzmann transport theory.
Instead, we will evaluate the spin boson life-time
$\tau_{\rm imp}$ using the
Green's function perturbative expansion in powers of $\delta J_{\rm imp}$.

There are two issues we must clarify before proceeding with this approach.
First, below the Kane-Fisher temperature the impurity scattering is very
strong and leads to the localization of 1D excitations.\cite{KaneFisher}
On the other hand, at the temperatures well above this scale
the impurity contribution can be analyzed perturbatively.
Thus, for the perturbation theory in powers of $\delta J_{\rm imp}$ to be
valid the temperature must be bigger than the Kane-Fisher temperature
\cite{KaneFisher}:
\begin{eqnarray}
\label{TKF}
T \gg T_{\rm KF} = J \left( \frac{\delta J_{\rm imp}}{J} \right)^
\frac{1}{1 - {\cal K}} = \frac{ \delta J_{\rm imp}^2}{J}.
\end{eqnarray}
Second, it is important to note that, generally, the life-time of an
excitation  is not equivalent to the transport relaxation time.
While the only processes which
violate the conservation of the
momentum contribute to the latter, any kind of
scattering shortens the former. However, since Eq. (\ref{Himp})
does not conserve momentum and since in any typical
scattering event the spin-boson momentum changes
drastically, there is no distinction between these two time scales in
our case. This justifies the use of our approach.

The correction to the single-boson Green's function from the impurity
scattering is evaluated as follows.
The lowest-order single impurity contribution to the Green's function is:
\begin{eqnarray}
&&{\cal D}_{k}(\tau) - {\cal D}_{0k}(\tau) \approx
\int
\left< \left( b^\dagger_k(\tau) + b^{\vphantom{\dagger}}_{-k}(\tau) \right)
\times\right.\\
&&\left.H_{\rm imp} (\tau') H_{\rm imp} (\tau'')
\left( b^{\vphantom{\dagger}}_{k'}(0) + b^\dagger_{-k'}(0)
\right)\right>_{\rm con-\atop nected}
d\tau ' d\tau ''.
\nonumber
\end{eqnarray}
The right hand side of this equation is proportional to the
self-energy. Since the impurity breaks the translational
invariance the self-energy depends on two momenta. As usual, the
translational invariance will be restored after averaging over the
random impurity positions. Since the perturbation $H_{\rm imp}$,
Eq. (\ref{Himp}), is an exponential in the bosonic field $\tilde
\Phi$ rather than a polynomial, the calculation of the Matsubara
average involves an effective summation of an infinite number of
terms, as schematically shown in Fig. \ref{imp_scatt}. Details of
such a procedure are given in Appendix \ref{app_B}. As a result of
our calculation the retarded self-energy is obtained though the
analytical continuation of the Matsubara self-energy:
\begin{eqnarray}
\label{Sigma1}
\Sigma^R_{k,\omega} \approx
- i\ \frac{n \ \delta J_{\rm imp}^2 \omega}{2a J |k|T} \ ,
\end{eqnarray}
with $n$ being the dimensionless impurity concentration. Note that
this self-energy is pure imaginary.
\begin{figure}[t]
\includegraphics[width=8cm]{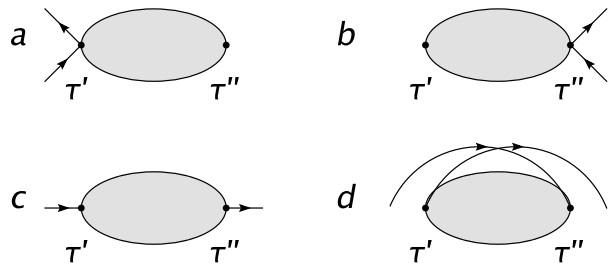}
\caption{Second-order diagrams for the spin-boson scattering on impurities.
Vertices represent interactions with impurities at times $\tau'$ and
$\tau''$. Shaded ellipses emphasize that there is a summation over virtual
states with any number of spin-boson excitations. This is due to the fact
that the interaction Hamiltonian is exponential in the bosonic field.
Formally, the ellipses correspond to
$\langle e^{i\sqrt{2\pi} \tilde \Phi (0, \tau')}
e^{i\sqrt{2\pi} \tilde \Phi (0, \tau'')}\rangle$.\cite{jaxodraw}}
\label{imp_scatt}
\end{figure}

Once the self-energy is found one can obtain the dressed
Green's function:
\begin{eqnarray}
{D}_{k,\omega} = \frac{D_{0k,\omega}}
{1 - { D}_{0k,\omega} \Sigma_{k,\omega}^R},
\end{eqnarray}
where
\begin{eqnarray}
D_{0k,\omega} = \frac{2\omega_k} {\omega_k^2 - \omega^2},
\end{eqnarray}
From here it is
possible to find the lifetime of the spin bosons
by  solving equations on the poles of the Green's function:
$1 = { D}_{0k,\omega} \Sigma_{k,\omega}^R$.
The yields the following expression for $\tau_{\rm imp}^{-1}$:
\begin{eqnarray}
\label{tau_imp}
\frac{1}{\tau_{\rm imp}}=  \frac{\Delta^2}{T},
\end{eqnarray}
where
\begin{eqnarray}
\Delta^2 \propto {n \delta J_{\rm imp}^2}.
\end{eqnarray}
The details of the derivation are given in Appendix B.
We would like to note that the impurity-induced relaxation rate
$\tau_{\rm imp}^{-1}\propto T^{-1}$ has been first obtained by Oshikawa and
Affleck, Ref. \onlinecite{Affleck}, in the context of the theory of
electron spin resonance in the S=1/2 quantum antiferromagnetic chains.

Unlike the spin-phonon scattering, the relaxation rate due to the disorder
is independent of $k$. The impurity scattering provides an effective
relaxation mechanism at low energies and thus removes
the divergence of the thermal conductivity in Eq. (\ref{k_diver}) by
effectively cutting off the low-energy spin bosons.

\begin{figure}[t]
\includegraphics[width=8.5cm]{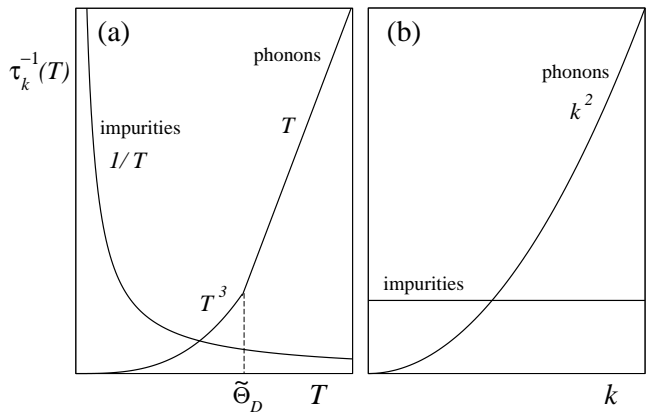}
\caption{The sketch of the spin-phonon and spin-impurity
relaxation rates $\tau_{sp}^{-1}$ and $\tau_{imp}^{-1}$ as a
function of (a) temperature, (b) momentum. In (a)
$\tilde\Theta_D=\Theta_D/4$.} \label{tau_pic}
\end{figure}

Our Fig. \ref{tau_pic} gives a qualitative comparison of the temperature
and $k$ dependencies of the spin-phonon and spin-impurity relaxation
rates $\tau_{sp}^{-1}$ and $\tau_{imp}^{-1}$.

\section{Thermal conductivity}
\label{kappa}

{\it Cumulative result.}\ \
Finally, having at hands the transport relaxation rates of the
spin-boson due to scattering on the phonons, Eq. (\ref{tau_tr000}), and on
impurities, Eq. (\ref{tau_imp}),
we are adequately equipped to calculate the thermal conductivity of
the spin chains.
The total relaxation time
 is given by:
\begin{eqnarray}
\label{tau_tot}
\frac{1}{\tau_{\rm tot}(k)} = \frac{1}{\tau_{\rm imp}} +
\frac{1}{\tau_{\rm sp}(k)}.
\end{eqnarray}
Then the solution of the Boltzmann equation which we had for the
phonon scattering (\ref{f1_ph}) should be modified to:
\begin{eqnarray}
f^1_k = \frac{v|k|\tau_{\rm tot}(k)}{T}
\frac{\partial f_k^0}{\partial k} \left(\partial_x T \right) \ .
\label{f1_tot}
\end{eqnarray}
As a result the thermal current is given by:
\begin{eqnarray}
J_E^s = \int_k v^2 k f^1_k =
({\partial_x T})\frac{v^3}{T} \int_k \frac{k|k|}
{\tau_{\rm imp}^{-1} + \tau_{\rm sp}^{-1}} \frac{\partial
  f_k^0}{\partial k} \ .\label{J1}
\end{eqnarray}
Thus, using the explicit expressions for the spin-phonon and spin-impurity
relaxation rates the following formula for the thermal conductivity
can be obtained:
\begin{eqnarray}
\label{kappa2}
\kappa_s (T) = \frac{v T^2}{\pi\Delta^2}
\int_0^{J/T}
\frac{x^2 \ dx}{4\sinh^2(x/2)}\ \frac{\alpha (T)}{x^2 + \alpha (T)}\ ,
\end{eqnarray}
where $x=vk/T$ and
\begin{eqnarray}
\label{a_T}
\alpha (T)=\frac{v^5 \Delta^2}{{\cal A} T^6\Gamma(\frac{\Theta_D}{T})}
\end{eqnarray}
with $\Gamma(z)$ defined in Eq. (\ref{gamma0}).
Since the temperature is always much smaller than $J$ one can safely
replace the upper limit in Eq. (\ref{kappa2}) by infinity.

Our Eq. (\ref{kappa2}) specifies a function with the following
properties. Depending on the value of $\Theta_D$ it either has a
single maximum at some $T = T_m$ (larger $\widetilde\Theta_D$) or
is a monotonic function (smaller $\widetilde\Theta_D$). In both
cases this function vanishes as $T^2$ at $T \ll T_m$ and saturates
at $T \gg T_m$. These qualitative features of $\kappa_s(T)$ can be
easily established. Without impurities ($\Delta = 0$) the integral
in Eq. (\ref{kappa2}) diverges at the lower limit. The impurity
scattering cuts off this divergence. When the temperature is small
the impurity scattering is much stronger than the typical
spin-phonon scattering, which corresponds to $\alpha (T)\gg 1$.
Thus, $\kappa_s\propto T^2$ at low temperatures. In the limit of
larger temperatures the impurity scattering is much weaker than
the spin-phonon one and $\alpha(T) \ll 1$.
If we assume that the temperature is still much less than the
Debye energy $T_m < T\ll\Theta_D$ then $\Gamma(\Theta_D/T)=1$.
Therefore, in this temperature range Eq. (\ref{kappa2}) yields
$\kappa_s\propto T^2\sqrt{\alpha (T)}\propto 1/T$. At yet higher
temperatures $T\agt {\widetilde\Theta_D}$ the temperature behavior
of the spin-phonon relaxation rate changes from $\propto T^3$ to
$\propto T$, which is reflected in the change of
$\Gamma(\Theta_D/T)$ from 1 to $\approx
({\widetilde\Theta_D}/T)^2$, see Eqs.
(\ref{tau_tr000})-(\ref{tau_tr0_a}). This gives $\kappa_s\propto
T^2\sqrt{\alpha (T)}\propto T^0$. Altogether, assuming
${\widetilde\Theta_D}$ is not too small:
\begin{eqnarray}
\label{kappa_q1}
\kappa_s(T)\propto\cases{
T^2   & for $T\ll T_m$ ,\cr
T^{-1}& for $T_m\ll T\ll {\widetilde\Theta_D}$ , \cr
T^0   & for $T\gg {\widetilde\Theta_D}$ . \cr}
\end{eqnarray}
where ${\widetilde\Theta_D}\approx \Theta_D/4$, as introduced in
Sec. \ref{phonons}. Note that the low-temperature,
impurity-controlled thermal conductivity in the generic Luttinger
liquids has been studied in Ref. \onlinecite{mus'yu} using the
memory-matrix formalism. Our $\propto T^2$ result matches the
result of that work for the Luttinger liquid parameter of the
spin-$\frac12$ chain ${\cal K}=1/2$. As one can see from Eq.
(\ref{kappa_q1}) our results do not support suggestion of
$\kappa_s\propto \exp(T^*/T)$ made in the experimental work, Ref.
\onlinecite{Sologubenko1}. Below we will discuss the origin of the
maximum and of the saturation value of $\kappa_s$, but first we
compare with the available experimental data.

\begin{figure}[t]
\includegraphics[width=8.5cm]{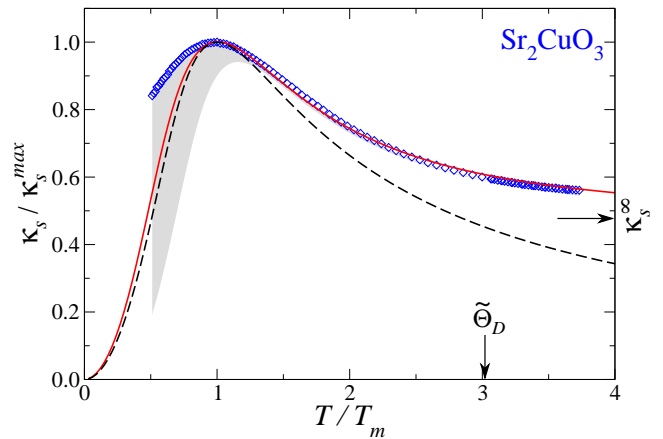}
\caption{Spin thermal conductivity $\kappa_s$
normalized to its maximal value vs
  reduced temperature $T/T_m$. Diamonds are the experimental data for
  Sr$_2$CuO$_3$, Ref. \onlinecite{Sologubenko1}. $T_m=79.4$K and
  $\kappa_s^{max}=36.7$Wm$^{-1}$K$^{-1}$ for this material. Shaded area
  schematically represents the range where the phonon background
  subtraction creates a large uncertainty in the data.\cite{Sologubenko1}
  Solid line is the results of this work, Eq. (\ref{kappa_u}), for
  $\Theta_D=11.6 T_m$. Arrows mark the saturation value $\kappa_s^\infty$
for the solid line
  at $T\gg {\widetilde \Theta_D}$ and the spin-phonon scattering crossover
  scale ${\widetilde \Theta_D}$. Dashed line corresponds to $\kappa_s(T)$, Eq.
(\ref{kappa_u}), for $\Theta_D \rightarrow \infty$. [These results
were recently presented in Ref. \onlinecite{us_PRL}.]}
\label{chain}
\end{figure}

{\it Comparison with experiments.}\ \
To compare our results with experimental data it is convenient to rewrite
the above expression for $\kappa_s(T)$, Eq. (\ref{kappa2}),
using the ``reduced'' temperature units $t=T/T_m$ with
$T_m$ which corresponds to the maximum in the
thermal conductivity:
\begin{eqnarray}
\frac{\kappa_s (t)}{\kappa_s^{max}} =
\frac{\gamma_1}{t^4 \Gamma(t)} \int_0^\infty
\frac{x^2 \ dx}{4\sinh^2(x/2)}\ \frac{1}{x^2 + \gamma_2/t^6\Gamma(t)}\ ,
\label{kappa_u}
\end{eqnarray}
where $\Gamma(t)\equiv \Gamma(\frac{\Theta_D}{T})$ from
Eq. (\ref{gamma0}), and
the constants $\gamma_1$ and $\gamma_2$ are chosen in such a way that
the temperature $T=T_m$ indeed corresponds
to a maximum in $\kappa_s$ and that
$\kappa_s(T_m)$ is indeed equal to $\kappa_s^{max}$.
For ${\widetilde\Theta_D}$ not too close to $T_m$ the constants are:
$\gamma_1=1.26$ and $\gamma_2=1.83$.
Such a choice of variables simply replaces a combination of
impurity- and lattice-related coupling constants
encoded in our $\Delta$ and ${\cal A}$, whose actual values are generally
unknown, by phenomenological constants $T_m$ and $\kappa_s^{max}$,
found from experiment. In fact, in the limit
$\Theta_D\rightarrow\infty$ this procedure
completely determines our $\kappa_s(t)$
given in Eq. (\ref{kappa_u}) since there are no adjustable parameters
left. The resulting $\kappa_s(T)$ for this limiting case
is shown in Fig. \ref{chain} by the dashed line.

Once we changed the variables to Eq. (\ref{kappa_u}) there is only
one parameter to adjust: $\Theta_D$. If we choose $\Theta_D = 11.6 T_m$ we fit
experimental data for the spin-chain compound Sr$_2$CuO$_3$ perfectly. For
this material $T_m = 79.4K$ and $\kappa_s^{max}=36.7$Wm$^{-1}$K$^{-1}$.
Figure \ref{chain} shows the comparison of our results and experimental data,
where the latter are shown by diamonds and the former is the solid line.
Shaded area highlights the region where, experimentally, the phonon background
subtraction creates a large uncertainty in the
data.\cite{Sologubenko1}

Note that at high
temperatures $T\gg {\widetilde\Theta_D}$ Eq. (\ref{kappa_u}) yields
the asymptotic value of the spin-boson thermal conductivity
$\kappa_s^\infty$:
\begin{eqnarray}
\frac{\kappa_s^\infty}{\kappa_s^{max}} =
\frac{\pi\gamma_1}{2\sqrt{\gamma_2}}\
\frac{T_m}{{\widetilde\Theta_D}}\approx 5.55 \frac{T_m}{\Theta_D}\ .
\label{kappa_as}
\end{eqnarray}
Therefore, when we fix $\Theta_D$ we fix $\kappa^\infty_s$ as well. Our
choice of $\Theta_D$ corresponds to  $\kappa_s^\infty = 0.48 \kappa_s^{max}$.
In Fig. \ref{chain} $\kappa^\infty_m$ is marked by an arrow.

We would like to emphasize that apart from $\kappa^{max}_s$ and $T_m$
there is only one adjustable parameter, $\Theta_D$, which
is used in our fitting procedure, and yet our theory yields an
excellent agreement with the experimental data over the whole
temperature range.

Note that the transition between  intermediate- and
high-temperature regimes for $\kappa_s$ is determined by the scale
${\widetilde\Theta_D}$, which separates different temperature behavior
for the spin-phonon scattering as discussed in Sec. \ref{phonons}.
It is this scale, not the Debye temperature itself, is a parameter of
our theory.
For Sr$_2$CuO$_3$ this scale is found to be
${\widetilde\Theta_D}\approx 3T_m=240$K, shown in Fig. \ref{chain} by
an arrow.

The value of the Debye temperature $\Theta_D$ that comes out from our fit
of thermal conductivity for Sr$_2$CuO$_3$ is
$\Theta_D=11.6 T_m \approx 900$K, which is higher than
the estimations of $\Theta_D$ for
this material obtained by a different method.\cite{Sologubenko1}
Since in our theory the Debye temperature is a characteristic energy scale for
the phonon bandwidth this may imply that the spin system is also
coupled to the
optical phonons. In fact, the typical width of the whole phonon
spectra  in the cuprates is about 900K.\cite{phonon_width}

To complete our comparison we provide here the asymptotic expressions
for $\kappa_s(T)$ in the low- and intermediate-temperature regimes.
At small
temperature $T_{\rm KF} \ll T \ll T_m$ the thermal conductivity is:
\begin{eqnarray}
\label{kappa_q2}
\frac{\kappa_s(T)}{\kappa_s^{max}}=
\frac{\pi^2\gamma_1}{3\gamma_2} \frac{T^2}{T_m^2}\approx
2.26 \frac{T^2}{T_m^2} .
\end{eqnarray}
In this regime the conductivity is controlled entirely by impurities.

When the temperature is higher the number of high-energy spin bosons
increases. The most important relaxation mechanism for these excitations is
the spin-phonon scattering. However, the low-energy spin bosons are
still scattered
mostly by impurities. Thus, at intermediate  temperature $T_m\ll T\ll
{\widetilde\Theta_D}$ both mechanisms contribute to the relaxation of
the thermal current:
\begin{eqnarray}
\label{kappa_q3}
\frac{\kappa_s(T)}{\kappa_s^{max}}=
\frac{\pi\gamma_1}{2\sqrt{\gamma_2}}\frac{T_m}{T} \approx 1.46
\frac{T_m}{T}.
\end{eqnarray}
These asymptotic formulas  can be tested experimentally, although  the
intermediate-temperature regime for Sr$_2$CuO$_3$ is not very well
pronounced because ${\widetilde\Theta_D}$ is only 3 times larger than
$T_m$. Such an intermediate asymptotic behavior
might be more relevant for some other systems
where the difference between $T_m$ and ${\widetilde\Theta_D}$ is larger.

For larger temperatures $T > \widetilde \Theta_D$ the thermal conductivity
saturates. Thus,
one of the most straightforward ways to verify our theory would be to
measure the thermal conductivity for higher temperatures and check if
the saturation really takes place as we predict.

\begin{figure}[t]
\includegraphics[width=8.5cm]{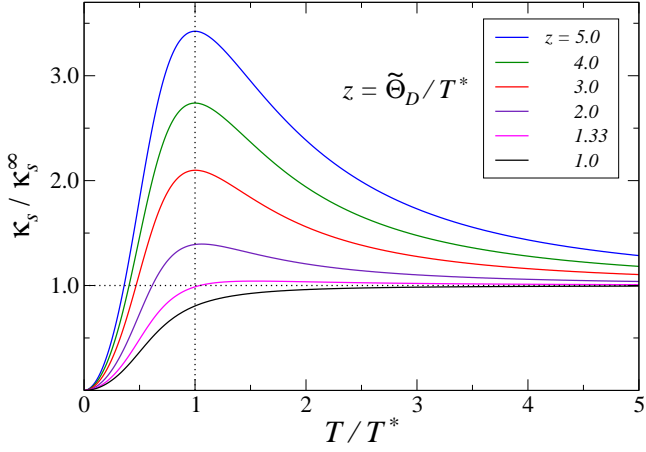}
\caption{Spin thermal conductivity $\kappa_s$
normalized to its value at $\infty$ vs
  reduced temperature $T/T^*$ for various values of
  $z=\widetilde\Theta_D/T^*$. ($z$ decreases from the top to the
  bottom curve).}
\label{chain_theta}
\end{figure}
{\it Different choices of $\Theta_D$.}\ \
One may wonder at this point how much variation in the temperature
dependence of $\kappa_s$ can be expected from the theory
given by Eq. (\ref{kappa2}) or (\ref{kappa_u}) if the parameters of
the model, such as $\Theta_D$, are varied. As is mentioned above there
are three independent parameters: phonon- and impurity-couplings
${\cal A}$ and $\Delta$, and the Debye temperature $\Theta_D$. To
generate a sequence of curves for the theoretical $\kappa_s$ v.s. $T$
the following procedure is convenient. Since at $T\gg\Theta_D$
thermal conductivity given by Eq. (\ref{kappa2}) saturates at a
constant value, it is convenient to normalize $\kappa_s(T)$ to its
value at $T=\infty$, i.e. $\kappa_s^\infty$, not to
$\kappa_s^{max}$. This is also because for $\widetilde\Theta_D\alt
1.25$ $\kappa_s(T)$ becomes a monotonic function and the determination
of $\kappa_s^{max}$ is ambiguous. Such a normalization reduces
the number of free parameters to two. It is natural to choose one of
those parameters to be a phonon crossover scale $\widetilde\Theta_D$
($\simeq \Theta_D/4$) and to define the other one as a reference
temperature $T^*$. Using Eq. (\ref{kappa_u}) with dimensionless
constants $\gamma_1$ and $\gamma_2$ related through:
\begin{eqnarray}
1=\frac{\pi\gamma_1}{2\sqrt{\gamma_2}}\
\frac{T^*}{{\widetilde\Theta_D}}\ ,
\label{g1_g2}
\end{eqnarray}
one can generate a sequence of curves for $\kappa_s/\kappa_s^\infty$
v.s. $t=T/T^*$ for various values of $z=\widetilde\Theta_D/T^*$ shown
in Fig. \ref{chain_theta}. Results for $z=$5, 4, 3, 2, 1.33, and 1
are shown ($z$ decreases from the top to the bottom curve).
The value of $\gamma_2=1.83$, same as in Fig. \ref{chain}, is used to
ensure that for $\widetilde\Theta_D\gg T^*$ (in reality, for
$\widetilde\Theta_D\agt 2T^*$) the reference temperature automatically
corresponds to the temperature of the maximum, i.e. $T^*\equiv
T_m$. Different choice of $\gamma_2$ would simply correspond to
rescaling of $T^*$ leaving the form of $\kappa_s(t)$ unchanged. Thus,
Fig. \ref{chain_theta} gives a description of possible variations in
the results of the theory given by Eq. (\ref{kappa2}) within
the experimentally relevant range of $\Theta_D/T^*$.

As is described above, in order to fix the remaining two parameters
and fit the experimental data one may associate the experimental value
of $T_m$ with the position of the theoretical maximum and then vary
$\widetilde\Theta_D$ to match $\kappa_s^{max}/\kappa_s^\infty$, for
example. A similar strategy has been used to obtain Fig. \ref{chain}
where the best fit closely corresponds to $z=3$ curve in
Fig. \ref{chain_theta}. If the experimental curve does not have a
maximum,\cite{MMM1} one can choose any other reasonably distinct
$\kappa_s(T')$ in order to fix the choice of theoretical parameters
$T^*$ and $\widetilde\Theta_D$.

{\it Origin of $T_m$.}\ \
The temperature $T_m$ at which $\kappa_s$ reaches its maximum can be
expressed through the original parameters of the problem as:
\begin{eqnarray}
T_m = \left( \frac{v^5 \Delta^2 }{ \gamma_2 {\cal A}} \right)^{1/6} \propto
v\left( \frac{n \delta J_{\rm imp}^2}{v{\cal A}} \right)^{1/6}\ .
\label{Tm}
\end{eqnarray}
In a similar fashion one can relate $\kappa_s^{max}$ to the
microscopic parameters as:
\begin{eqnarray}
\kappa_s^{max} =\frac{v^6}{\gamma_1\pi{\cal A} T_m^4} \propto
\frac{v^2}{\cal A} \left(
\frac{v{\cal A}}{n \delta J^2_{\rm imp}} \right)^{2/3}. \label{kmax}
\end{eqnarray}
The physical meaning of $T_m$ is the following. It is possible to
define the ``thermal'' spin-phonon relaxation rate:
\begin{eqnarray}
\label{tau_th}
\frac1{\tau^T_{sp}} = \frac{1}{\tau_{sp}(k)}\bigg|_{k = k_T} = \frac{{\cal
A} T^5}{v^5},
\end{eqnarray}
where $k_T=T/v$.
By its definition $\tau^T_{sp}$ is independent of $k$. It
characterizes rate of the spin-phonon collisions of the thermal spin bosons
at a given temperature. The scale $T_m$ emerges naturally as a solution of the
equation:
\begin{eqnarray}
\tau^T_{sp}= \tau_{\rm imp}.
\end{eqnarray}
In other words, when $T \propto T_m$ the thermal spin-phonon and
spin-impurity collision rates are the same.
At the temperature higher (lower) that $T_m$ the spin-phonon
(impurity) scattering dominates. Note that in the
low-temperature regime we assume that the temperature is still much
larger than the Kane-Fisher temperature $T_{KF}$.

{\it Origin of $\kappa_s^\infty$.} \ \
Our work predicts a remarkable behavior of the thermal conductivity at
the temperatures higher than the scale related to the Debye energy:
saturation at a constant value, Eq. (\ref{kappa_as}).
A parallel can be drawn with the thermal conductivity of metals, where
the electronic part of $\kappa$ saturates at the temperatures
above the Debye energy.\cite{Berman} The parallel can be suspected to
be even deeper
because of the presence of the large energy scales in both problems,
Fermi energy $E_F\gg T$ in the case of metals and $J\gg T$ in our case.

A closer examination, however, reveals a difference between the two
cases. In metals, the thermal conductivity temperature behavior is
defined by the electronic specific
heat $C_e(T)$ and the electron-phonon relaxation
time $\tau_{ep}(T)$ through the quasiclassical relation:
$\kappa_e(T)=C_e(T)v_F^2\tau_{ep}(T)/3$. Note that it is the
thermalized electrons which
contribute most substantially to the transport. Specific heat is
linear in $T$, while at $T\agt \Theta_D$ the relaxation
time is inversely proportional to the temperature
 due to the thermal population of phonons,
$\tau_{ep}(T)\propto 1/T$. Altogether,
this renders $\kappa_e=const$.

In our case the excitations are spin bosons, not
fermions at large momentum $k_F$.
Therefore, in contrast, in the phonon-scattering
dominated regime $T\gg T_m$ the spin-boson relaxation rate is strongly
$k$-dependent and the major contribution to the thermal
current comes from the long-wavelength bosons with $k\ll T/v$. Thus,
the impurity-scattering continues to contribute to the temperature dependence
of the thermal conductivity at higher temperatures.
Therefore, in the case of spin chains the
saturation of $\kappa_s$ is a result of a non-trivial combination of:
particular temperature- and $k$-dependencies of the
impurity and phonon relaxation rates and the 1D density of states of
the spin-bosons.

{\it Further predictions.}  \ \
Our equations (\ref{kappa_as}), (\ref{Tm}) and (\ref{kmax}) allow us
to formulate several predictions of our model for the impurity
concentration dependence of several quantities
that can be verified experimentally.

First, we see that the temperature $T_{m}$ at which $\kappa_s(T)$ reaches
its maximum scales as $n^{1/6}$.
Such a weak dependence means that for a particular material
the maximum in $\kappa_s$ should be
around the same temperature for a broad range of disorder
concentration.

Second, the maximum value of the thermal
conductivity $\kappa_s^{max} = \kappa_s(T_{m})$ scales as $n^{-2/3}$,
Eq. (\ref{kmax}).
Third,
the saturation value $\kappa_s^\infty$ has yet different concentration
dependence $n^{-1/2}$. Such behaviors can be looked for in the
materials with the isotope substitution for either magnetic ions
(Cu$^{2+}$) or surrounding ions as they all will lead to the local
modification of the superexchange constant. If such dependencies are
observed this will provide strong support to our theory.

Another interesting experimental suggestion is based on the specifics of the
spin-phonon scattering which we have found in Sec. \ref{phonons}. Namely,
the most important scattering of spin bosons in the phonon-dominated regime
is due to phonons whose  momentum is almost normal to the direction of the
spin chains. The characteristic energy of such phonons is $T$. One can think
of the following experimental setup: the heat current is directed along the
chains while the thermal phonon pulse is induced perpendicular to the heat
current (e.g., by a short laser pulse). This should lead to a sizable
suppression of the spin-boson thermal current due to additional scattering
by the extra phonons. The increase of the phonon heat current due to
transverse pulse will be either negligible or can be accounted for by
comparing to the results of the same experiment with the heat current and
heat pulse both directed perpendicular to the chains.

{\it Consistency check.} \ \
In considering the spin-boson relaxation due to phonons we emphasized
the importance of the two-stage, bottle-neck process, in which the
spin boson scatters on the phonon via a ``normal'' process, while the
momentum relaxation via an Umklapp process is taken care of by the
phonon bath.
Having carried out a comparison  with
experiments we can now verify the consistency of our results with the
initial assumption. Namely, we can now check whether the
characteristic spin-phonon relaxation time $\tau_{sp}$ is indeed much
longer than the characteristic phonon-phonon relaxation time $\tau_{pp}$.

Since below $T_m$ the scattering is impurity-dominated we need to
estimate the relaxation times for $T\geq T_m$ only. We have already
introduced the ``thermal'' relaxation time for spin bosons $\tau^T_{sp}$
in Eq. (\ref{tau_th}). At $T=T_m$ one can express such relaxation time
through $\kappa_s^{max}$, Eq. (\ref{kmax}), as:
\begin{eqnarray}
\tau^T_{sp}=\frac{\gamma_1\pi\kappa_s^{max}}{v T_m}  \
. \label{tau_th1}
\end{eqnarray}
The phonon relaxation time can be estimated from the quasiclassical
expression for the phonon thermal conductivity:
$\kappa_{ph}=c^2C(T_m){\widetilde{\tau}}_{pp}/3$, where $C(T_m)$ is
the lattice specific heat and ${\widetilde{\tau}}_{pp}$ is the average
transport relaxation time of a phonon due to phonon-phonon scattering.

Using these expressions we write the ratio of the thermal spin-phonon
and the average phonon-phonon relaxation times as:
\begin{eqnarray}
\frac{\tau^T_{sp}}{\widetilde{\tau}_{pp}}=\gamma_1
\left(\frac{C(T_m)}3\right)
\left(\frac{\kappa_s^{max}}{\kappa_{ph}}\right)
\left(\frac{c}{v}\right)\left(\frac{\Theta_D}{T_m}\right)
\
, \label{tau_2}
\end{eqnarray}
where the specific heat $C(T_m)$ can be found as $\min\{3; 12\pi^2
(T_m/\Theta_D)^3/5\}$.\cite{Ziman}
One can see that the smallness of $c/v$ in
(\ref{tau_2}) is canceled by the largeness of $\Theta_D/T_m$.
For  Sr$_2$CuO$_3$ in the region around $T_m$ the spin-boson and
phonon parts of the thermal conductivity are of the same order.
Also, $C(T_m)\approx 2$.
Taking the actual values of $\kappa_s^{max}$, $\kappa_{ph}$, $T_m$,
$J$, etc. for Sr$_2$CuO$_3$ we find that:
$\tau^T_{sp}/\widetilde{\tau}_{pp}\simeq 0.5\div 1$.

It appears, superficially, that our assumption is not valid. However,
at $T>T_m$ the typical momentum $\widetilde{k}$
of a spin boson contributing to the thermal transport is
much smaller than the ``thermal'' momentum
$k_T=T/v$: $\widetilde{k}/k_T\approx (T_m/T)^3$.
Thus, the typical relaxation time is much larger than the thermal one
$\tau_{sp}\approx \tau^T_{sp} \cdot (k_T/\widetilde{k})^2$.
We also recall here that the phonons contributing to our two-stage process
most effectively are the thermalized phonon with the momentum $P_T\sim
T/c$ as shown in Sec. \ref{phonons}. Therefore,
$\tau_{pp}^T$ of such a phonon is much shorter than
the average transport relaxation time $\widetilde{\tau}_{pp}$ which we
estimated above, $\tau_{pp}^T\ll \widetilde{\tau}_{pp}$.
Thus, the characteristic spin-phonon and phonon-phonon times
involved in the two-stage process obey:
\begin{eqnarray}
\frac{\tau_{sp}}{\tau_{pp}^T}\gg
\frac{\tau^T_{sp}}{\widetilde{\tau}_{pp}}\approx 1\ , \label{tau_3}
\end{eqnarray}
due to both $\tau_{sp}\gg \tau^T_{sp}$ and $\tau_{pp}^T\ll
\widetilde{\tau}_{pp}$.
This proves the validity of our approach.

We also estimated the Kane-Fisher temperature for Sr$_2$CuO$_3$ from
\begin{eqnarray}
\label{TKF0}
\kappa_s^{max}\approx \frac{aT_{m}^2}{nT_{KF}} \ ,
\end{eqnarray}
which follows from
Eqs. (\ref{TKF}), (\ref{Tm}), and (\ref{kmax}). Using experimental values
for $\kappa_s^{max}$ and $T_m$ one obtains that
\begin{eqnarray}
\label{TKF1}
\frac{T_{KF}}{T_m}\approx 10^{-3} n^{-1} \ ,
\end{eqnarray}
which means that our assumption for the impurity-controlled regime
$T_{KF}\ll T_m$ is valid unless the impurity concentration $n$ is
less than 0.1\%. We are not aware of the actual level of disorder
in Sr$_2$CuO$_3$, but such a high purity is unlikely. Therefore,
this assumption is valid too.

With the help of our asymptotic formula (\ref{kmax}) one can
estimate the spin-coupling constant using experimental values of
$\kappa_s^{max}$ and $T_m$. This estimate leads to an unreasonably
high value of ${\cal A}/Ja^5\sim 10^4$. First, a large value of
the spin-phonon coupling constant can be due to an exponential
dependence of the superexchange on the interatomic distance, as it
is generated by the virtual tunnelling of electrons. Second, as
one can see from Eqs. (\ref{Tm}) and (\ref{kmax}) the expression
for the spin-phonon coupling contains high powers of both $J$ and
$T_m$. This creates an extra sensitivity of the result of our
estimate to the unknown dimensionless constants, which our theory
might be unable to capture.\cite{comment}

{\it Other materials.} \ \
Other materials of the cuprate family in which the anomalous thermal
transport has been observed include the
zig-zag chain\cite{Sologubenko1} and spin-ladder materials,\cite{Hess} and
another spin-chain material,
BaCu$_2$Si$_2$O$_7$, with much smaller
superexchange constant.\cite{Sologubenko2}

The most straightforward case for applying our theory seems to be
the case of SrCuO$_2$ where the spin excitation spectrum
remains gapless because of the frustrated coupling between
the chains. Since the inter-chain coupling is of order
$J'/J\sim 0.1$ one may expect that while the low- and intermediate-$T$
behavior of $\kappa_s$ should be the same as in the case of Sr$_2$CuO$_3$,
the high-$T$ part might be
different due to an additional scattering mechanism
from the inter-chain interaction. However, as we show in
Section \ref{meanfp}, the high-$T$ regime is fully dominated by the
long-wavelength spin excitations, which means that $J'$ should remain
irrelevant and the behavior of $\kappa_s(T)$ for SrCuO$_2$
should be very similar to the ``simple'' chain material Sr$_2$CuO$_3$
at all temperatures.
Nevertheless, we were unable to fit the experimental data
for SrCuO$_2$\cite{Sologubenko1} as successfully as for Sr$_2$CuO$_3$.
On the other hand, as we learned from a very recent work
Ref. \onlinecite{MMM1} there might be an issue with the subtraction
background for these materials. The
thermal conductivity of the 5\% Ca-doped SrCuO$_2$ does exhibit a
saturation at high-temperatures, in an excellent agreement with our
theory, but the peak at lower temperatures is much less pronounced.
We therefore refrain from an explicit comparison of our theory to
SrCuO$_2$ until the experimental issues are settled.

In the case of (La,Ca,Sr)$_{14}$Cu$_{24}$O$_{41}$ materials the
excitations are gapped. Therefore, the results of our work cannot be
straightforwardly applied to them because the gap, not the impurity
scattering, controls the  low-energy cut-off scale.
Thus, one needs to reconsider the
problem of the thermal transport for the gapped spin systems.
On the other hand, the spin-phonon
scattering part of such a consideration may stay very similar to our
case, especially for
Ca$_9$La$_5$Cu$_{24}$O$_{41}$. This latter
compound presumably contains a lot of 3D disorder (not related to the
spin subsystem) due to a random distribution of Ca and La ions,\cite{Hess}
which might help to facilitate the fast phonon momentum relaxation and
ensure that $\tau_{pp}\ll \tau_{sp}$ at high temperatures. We would
like to note that another very recent experimental work,
Ref. \onlinecite{MMM2}, has discussed the role of the spin-phonon
coupling in the gapped systems using a very different framework.

The case of another spin-chain material,
BaCu$_2$Si$_2$O$_7$, is also different because of the lack
of the spin and lattice energy scales separation. In this
compound the spin boson velocity
is of the same order as the sound velocity,
$v\approx c$.\cite{Sologubenko2}
Therefore, the kinematic considerations of our work will not be
applicable for the collision integral in this system.  In the
temperature regime $T\sim J$, easily achievable for this material,
the thermal transport will be influenced
by various other relaxation processes, which are negligible in our
case.

Altogether, the differences in the underlying spin models for some of
these materials  will, most certainly, lead to  different
temperature behaviors of $\kappa_s$.
Therefore, a detailed microscopic study in these systems along the lines of
our work is necessary.

\section{mean free path}
\label{meanfp}

Having described successfully the temperature dependence of the
thermal conductivity
in Sr$_2$CuO$_3$ we would like to discuss the length scales that characterize
scattering in this system. We also would like to address the question
of whether
the concept of the mean free path can be applied to describe thermal
conductivity by spin bosons.
More specifically, we would like to ask whether the thermal
conductivity of spin bosons can be written in a simplified form
(sometimes referred to as the simple kinetic equation):
\begin{eqnarray}
\label{kinetic}
\kappa_s(T)=v C_s(T)\bar{\ell}(T)\ ,
\end{eqnarray}
and if yes in what regime.
Here $C_s(T)$ is the specific heat of the spin chains and
$\bar{\ell}(T)$ is the
mean free path.

In the end, we also verify that the wavelength of a typical excitation
contributing to the thermal
current in Sr$_2$CuO$_3$ is shorter than its mean free path.
This condition is needed for the applicability of the Boltzmann
equation formalism (quasiclassical approximation).

The ``total'' relaxation time of the spin boson with the
wave-vector $k$ is defined in Eq. (\ref{tau_tot}).
Thus, one can introduce corresponding
$k$- and $T$-dependent length scale:
\begin{eqnarray}
\label{ell_kT}
\ell(k,T)=\frac{v}{\tau_{\rm imp}^{-1}(T)+\tau_{\rm sp}^{-1}(k,T)}
\ .
\end{eqnarray}
In order to be clear on terminology we would like to refer to this
$k$-dependent length scale as to the ``$k$-dependent mean free path''
 of the spin boson. At the same time the notion
``mean free path'' will be reserved for
 an averaged, only $T$-dependent, length scale.

{\it Cut-off length.}\ \
Spin-phonon scattering rate in Eq. (\ref{ell_kT}) depends on $k$ as
$\tau^{-1}_{\rm sp}\propto k^2$
while impurity scattering rate is $k$-independent and plays the role
of a cut-off scale.
Therefore, it is natural to introduce the ``cut-off'' length as:
\begin{eqnarray}
\label{ell_coff}
\ell_{\rm imp}(T)\equiv \ell(0,T)=\frac{vT}{\Delta^2}\ ,
\end{eqnarray}
using explicit expression for $\tau^{-1}_{\rm imp}$, Eq. (\ref{tau_imp}).
It is interesting to find out the absolute value of such a length for
the real system like
Sr$_2$CuO$_3$. Since all the impurity- and phonon-related parameters,
$\Delta$, ${\cal A}$, and ${\widetilde\Theta}_D$, have been related to
the phenomenological
constants $T_m$($\simeq 80K$) and $\kappa_s^{max}$($\simeq 37$ W
m$^{-1}$K$^{-1}$),
one can re-express $\Delta$ through them and obtain:
\begin{eqnarray}
\label{ell_coff1}
\ell_{\rm imp}(T)=\frac{\pi\gamma_1}{\gamma_2}\
\frac{\hbar\kappa_{s}^{max}\ b\ c}{k_B^2T_m\sqrt{2}} \
\frac{T}{T_m} \simeq 300a\ \frac{T}{T_m}\ ,
\end{eqnarray}
where $a=3.9$\AA\ and $b=3.5$\AA\ and $c=12.7$\AA\ are
the lattice constants along and perpendicular to the chains in
Sr$_2$CuO$_3$, respectively.\cite{Johnston}
Such a large value of $\ell_{\rm imp}$ ($\approx 1200$\AA\ at $T=T_m$)
is in agreement with
the estimates  for the mean free
path made in the experimental works.\cite{Sologubenko1,Sologubenko_1a}
Note that the cut-off length {\it grows} as the temperature
increases. This is
a consequence of the impurity
scattering in 1D spin-chains which is non-perturbative in the number
of spin bosons.
\begin{figure}[t]
\includegraphics[width=8cm]{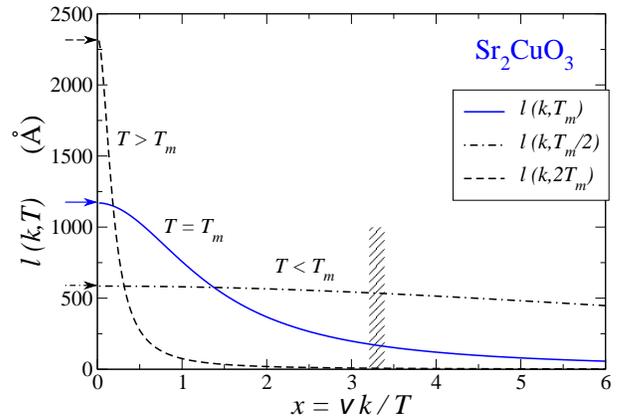}
\caption{$k$-dependent mean free path $\ell(k,T)$
v.s. $x=vk/T$ for three
representative temperatures: $T=T_m$ (solid), $T=T_m/2$ (dot-dashed),
and $T=2T_m$ (dashed). Arrows mark the cut-off length $\ell_{\rm
  imp}(T)$ for each temperature. An approximate  boundary of
the bosonic thermal population is marked by the shaded area.}
\label{ell_x}
\end{figure}

{\it $k$-dependent mean free path. }\ \
We now express the $k$-dependent mean free path
(\ref{ell_kT}) through the cut-off length (\ref{ell_coff})
using the explicit form of the
spin-phonon scattering rate, Eq. (\ref{tau_tr000}):
\begin{eqnarray}
\label{ell_xT}
\frac{\ell(k,T)}{\ell_{\rm
    imp}(T)}=\left[1+\frac{1}{\gamma_2}\left(\frac{vk}{T}\right)^2
\left(\frac{T}{T_m}\right)^6
\Gamma\left(\frac{{\widetilde\Theta}_D}{T}\right)\right]^{-1},
\end{eqnarray}
where $\Gamma({\widetilde\Theta}_D/T)$ is the auxiliary crossover
function introduced in Eq. (\ref{gamma0}), which is related to the
Debye temperature: $\Gamma\approx 1$ for
$T<{\widetilde\Theta}_D$ and
$\Gamma\approx ({\widetilde\Theta}_D/T)^2$ for $T>{\widetilde\Theta}_D$.
Our Fig. \ref{ell_x} shows the $k$-dependent mean free path $\ell(k,T)$
v.s. $x=vk/T$ for three
representative temperatures: $T=T_m$ (solid), $T=T_m/2$ (dot-dashed),
and $T=2T_m$ (dashed). From Eq.
(\ref{ell_xT}) one can see that $\ell(x,T)$  is a Lorentzian with the
height given by the cut-off length
$\ell_{\rm imp}(T)$ (marked by arrows in Fig. \ref{ell_x}) and a
temperature-dependent width.
This width depends on $T$ quite strongly: $\propto T^{-3}$ for
$T<{\widetilde\Theta}_D$ and
$\propto T^{-2}$ for $T>{\widetilde\Theta}_D$.

The spin boson energy normalized by to the temperature $x=vk/T$ is a natural
variable to describe thermal processes.
The values of $x$ which are of significance for such processes are
$\alt 3$ because of the bosonic thermal population;
this boundary is marked by
the shaded area in Fig. \ref{ell_x}. One can see that the
$k$-dependent mean free path is essentially constant for $x<3$
in the impurity-dominated regime $T=T_m/2$,
while it depends
on $x$ very strongly in the
phonon-dominated regime $T=2T_m$.

{\it Impurity-dominated regime.}\ \ Already from this observation one
can see that in the
impurity-dominated regime all the spin bosons involved in the thermal
transport are characterized by approximately the same
mean free path
$\ell(k,T)\approx\ell_{\rm imp}(T)$, Fig. \ref{ell_x}
dash-dotted line, and thus the mean free path
concept readily applies here. Namely, one can re-write our expression
for the thermal conductivity, Eq. (\ref{kappa2}), somewhat differently, using
the $k$-dependent mean free path from Eqs. (\ref{ell_kT}) and (\ref{ell_xT}):
\begin{eqnarray}
\label{kappa_ell}
\kappa_s(T)=\frac{T}{\pi}\int_0^\infty \frac{x^2
  e^x}{(e^x-1)^2}\ \ell(x,T)\ dx\ .
\end{eqnarray}
Since $\ell(x,T)$ is $x$-independent in the impurity-dominated regime,
this expression can be reduced to the form (\ref{kinetic})
\begin{eqnarray}
\label{k_qcl}
\kappa_s(T)\ =\ v\ C_s(T)\ \ell_{\rm imp}(T),
\end{eqnarray}
with the mean free path equal to the cut-off length
$\bar{\ell}(T)\equiv\ell_{\rm imp}(T)$, Eq. (\ref{ell_coff1}), and
the specific heat given by:
\begin{eqnarray}
\label{C_s}
C_s(T)=\int_k v|k|\frac{\partial f_k}{\partial T}=\frac{T}{\pi v}
\int_0^\infty \frac{x^2 e^x dx}{(e^x-1)^2}= \frac23\frac{T}{Ja} \ ,
\end{eqnarray}
where $v=\pi J/2$ and the well-known value of the integral
($=\pi^2/3$) have been used. Thus, our result for the mean free
path in the low-$T$ regime ($T<T_m$) is  $\ell\propto T$, which
disagrees with experimental works where often a constant mean free
path at low temperatures is
reported.\cite{Sologubenko_1a,Sologubenko2} However, this is the
regime where experimental data for the spin part of the thermal
conductivity are unreliable due to large phonon background.

\begin{figure}[t]
\includegraphics[width=8cm]{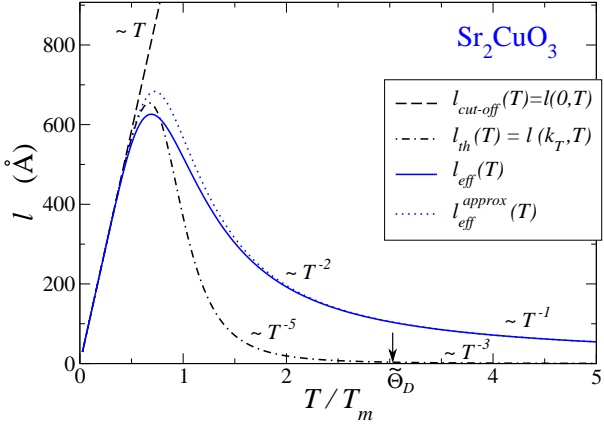}
\caption{The temperature dependence of the cut-off length $\ell_{\rm imp}$
($\propto T$, dashed) and the ``thermal'' length
$\ell_{th}(T)=\ell(T/v,T)$ (dashed-dotted).
In the phonon-dominated regime $\ell_{th}(T)\propto T^{-5}$ for
$T<\widetilde\Theta_D$ and $\propto T^{-3}$ for
$T>\widetilde\Theta_D$). The ``effective'' mean free path
$\ell_{eff}(T)$, Eq. (\ref{ell_eff}),
and the approximate  ``effective'' mean free path
$\ell^{app}_{eff}(T)$, Eq. (\ref{ell_eff_app}),
are also shown (solid and dotted, respectively).
}
\label{ell_T}
\end{figure}
{\it Phonon-dominated regime.}\ \ One can see from Fig.
\ref{ell_x} that the situation for $\ell(x,T)$ is quite different
in the phonon-dominated regime. While the cut-off length grows as
the temperature increases, the range of the energies of the spin
bosons effectively participating in the thermal transport shrinks.
In fact, it is a very unusual feature of the problem: as the
temperature increases the thermal current becomes dominated by the
excitations with longer and longer wavelengths. Therefore, our
approximations related to spin excitations, such as the use of
bosonization language, become better justified for higher
temperatures.

Using Eq. (\ref{ell_xT}) one can
define the  ``typical'' momentum of the spin boson $\widetilde k$,
corresponding to the width of the Lorentzian $\ell(\widetilde
k,T)=\ell_{\rm imp}/2$:
\begin{eqnarray}
\label{k_typ}
&&\widetilde k \approx
\frac{T_m}{Ja}\left(\frac{T_m}{T}\right)^2\approx
\frac{1}{20a}
\left(\frac{T_m}{T}\right)^2
 \
\mbox{for}\ \ \ T<\widetilde\Theta_D,
\nonumber\\
&&
\widetilde k \approx
\frac{T_m}{Ja}\frac{T_m}{\widetilde\Theta_D}\frac{T_m}{T}
\approx\frac{1}{60a}\frac{T_m}{T}
\ \ \ \ \ \ \ \ \
\mbox{for}\ \ \ T>\widetilde\Theta_D,
\end{eqnarray}
where the values $J/T_m=20$ and $\widetilde\Theta_D/T_m=3$ for
Sr$_2$CuO$_3$ were used.

Evidently, at $T>T_m$ one cannot possibly
assume that all the spin-boson excitations are
characterized by the same $k$-dependent mean free path. Therefore, one
cannot reduce
Eq. (\ref{kappa_ell}) for $\kappa_s(T)$ to a simplified
form, Eq. (\ref{kinetic}). In
other words, the mean free path concept cannot be, strictly speaking,
applied to the phonon-dominated regime of the problem. This is further
demonstrated in our Fig. \ref{ell_T} which shows the temperature
dependence of the cut-off length $\ell_{\rm imp}$
($\propto T$, dashed) and the ``thermal'' length
$\ell_{th}(T)=\ell(T/v,T)$ ($k$-dependent mean free path for the
excitation with
the energy $=T$, dashed-dotted). The latter length, while
indistinguishable from the cut-off length in the impurity-dominated
regime $T<T_m$, dies off very quickly in the phonon-dominated regime $T>T_m$
(as $\propto T^{-5}$ for
$T<\widetilde\Theta_D$ and then as $\propto T^{-3}$ for
$T>\widetilde\Theta_D$). This is simply to reiterate our point that
the spin bosons are characterized by strongly $k$-dependent
mean free paths for $T>T_m$.

{\it ``Effective'' mean free path.}\ \
Nevertheless, one may want to insist on using the simplified
equation (\ref{kinetic})
in order to extract some ``effective'' mean free path
$\ell_{eff}(T)$
characterizing the system, an approach used in the experimental
works,\cite{Sologubenko1,Sologubenko_1a}
\begin{eqnarray}
\label{kappa_eff}
\kappa_s(T)\ {\buildrel\rm def\over=}\ v\ C_s(T)\ \ell_{eff}(T).
\end{eqnarray}
Using the explicit form of $C_s(T)$ obtained above (\ref{C_s})
the ``effective'' mean free path  $\ell_{eff}(T)$ can be written  as:
\begin{eqnarray}
\label{ell_eff}
\ell_{eff}(T)\ {\buildrel\rm def\over=}\ \frac{3}{\pi^2}\int_0^\infty
\frac{x^2 e^x}{(e^x-1)^2}\ \ell(x,T)\ dx\ .
\end{eqnarray}
This integral can be taken numerically and $\ell_{eff}(T)$
is plotted in Fig. \ref{ell_T} (solid line).
Note again that all three lengths, $\ell_{\rm imp}(T)$,
$\ell_{th}(T)$, and $\ell_{eff}(T)$ are indistinguishable for $T<T_m$
while they are very different for $T>T_m$.

Recalling that the height of the Lorentzian
$\ell(x,T)$ grows $\propto T$  while
the width is decreasing $\propto T^{-3}$ for
$T<{\widetilde\Theta}_D$ and
$\propto T^{-2}$ for $T>{\widetilde\Theta}_D$ one can immediately
conclude that $\ell_{eff}(T)$ should decay as $\propto T^{-2}$ for
$T<{\widetilde\Theta}_D$ and
$\propto T^{-1}$ for $T>{\widetilde\Theta}_D$ as marked in
Fig. \ref{ell_T}. Since $C_s(T)\propto T$, this yields
$\kappa_s\propto const$ behavior for $T>{\widetilde\Theta}_D$.

Using the fact that the width of $\ell(x,T)$ is strongly $T$-dependent
one can obtain an approximate expression for
$\ell_{eff}(T)$ in Eq. (\ref{ell_eff}) which is valid in both
asymptotic limits $T\ll T_m$ and $T\gg T_m$:
\begin{eqnarray}
\label{ell_eff_app}
\ell^{app}_{eff}(T)\ =\
\frac{\ell_{\rm imp}(T_m)}
{Bt^2\sqrt{\Gamma(t)}}\arctan(Bt^3\sqrt{\Gamma(t)}),
\end{eqnarray}
where $t=T/T_m$ and constant $B=2.5$. It is shown in Fig. \ref{ell_T}
(dotted line). Using explicit form of $\Gamma(\widetilde\Theta_D/T)$
one can obtain all the correct asymptotic behaviors of
$\ell_{eff}(T)$ from it.

\begin{figure}[t]
\includegraphics[width=8cm]{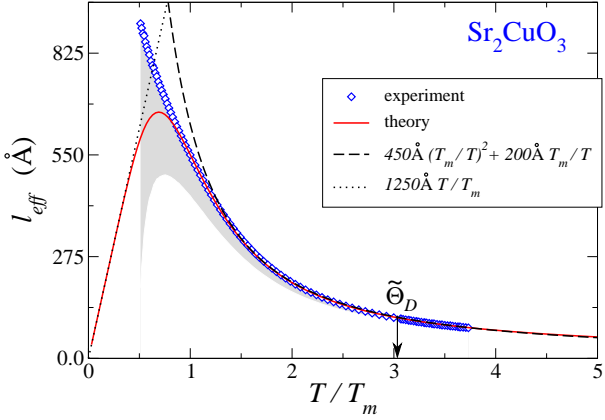}
\caption{The ``effective'' mean free path,
Eq. (\ref{ell_eff}), compared
to the experimental results of
Refs. \onlinecite{Sologubenko,Sologubenko1}.
Asymptotic fits $\propto
T$ for $T<T_m$ and $\propto 1/T+C/T^2$ for $T>T_m$ are shown by the
dashed lines. Shaded area schematically
represents the experimental error bars due to the phonon subtraction.
}
\label{ell_T1}
\end{figure}
Our Fig. \ref{ell_T1} shows our ``effective'' mean free path,
Eq. (\ref{ell_eff}),  compared
to the same quantity derived in
Refs. \onlinecite{Sologubenko1,Sologubenko_1a}
from the experimentally measured $\kappa_s(T)$
using Eq. (\ref{kappa_eff}). Asymptotic fits $\propto
T$ for $T<T_m$ and $\propto 1/T+C/T^2$ for $T>T_m$ are also shown. Shaded area
schematically
represents the experimental error bars due to the phonon subtraction.

{\it Saturation in $\kappa_s$ and parallel with metals.}\ \
We would like to return to the discussion of the origin of saturation in
$\kappa_s(T)$ at $T>\widetilde\Theta_D$. One may be surprised, once
again, by the seemingly identical features of the high-$T$ thermal
conductivity in both the spin chains
and metals: (i) in both cases there is  large scale ($E_F$ and $J$,
respectively) which yields ``fast'' excitations, (ii) in both cases
specific heat $\propto T$ up to very high temperatures, (iii) for the
saturation regime the dominant scattering is due to phonons in both
cases, (iv)  mean free path for metals and ``effective'' mean free
path for spin bosons are $\propto 1/T$ in this regime.

It is very tempting to conclude that in the case of the spin chains
the $1/T$ behavior in the ``effective'' mean free path is  due
to phonon population $n_{ph}\propto T$ as in the case of metals.
However, it is much more subtle in this case. As we discussed above
and in previous Section, thermal transport in spin chains
relies on the long-wavelength bosons, even more so for the high
temperatures (see Fig. \ref{ell_x}), and the $T$-dependence of the
impurity scattering remains important. Simply put: if the
impurity scattering would provide $\ell_{\rm imp}$ v.s. $T$ different
from $\propto T$, the $T$-dependence of the $\ell_{eff}$ would be
different from $\propto 1/T$ and there would be no saturation in
$\kappa_s(T)$. In fact, this seems to be the case for the related
frustrated ladder
(zig-zag chain) material SrCuO$_2$, which has gapless excitations as in
spin chains, but its thermal conductivity does not seem to saturate at
higher $T$.\cite{Sologubenko1}

Once again, it is an interesting and unusual situation in spin chains:
it is always the long-wavelength range of excitations which is
crucial for the thermal transport, temperature increase only
enhances the phonon scattering and shrinks the relevant energy
region for spin bosons.
Thus, our results are universal and do not depend on the
manner in which bosonization or other approximations break down at
higher energies.

{\it Quasiclassical approximation.}\ \
It is instructive to verify the validity of the quasiclassical
approximation which is implicitly used when utilizing Boltzmann equation
for the lower-order scatterings. One needs to show  for the
``typical'' spin boson that its wavelength $\widetilde \lambda$ is
shorter than its $k$-dependent mean free path $\widetilde \lambda <
\widetilde\ell$.

{\it $T>T_m$ case.} \ \
Consider $T>T_m$ first. We have already defined the ``typical''
momentum for this case in Eq. (\ref{k_typ}). Using $J/T_m=20$ and
  $\widetilde\Theta_D/T_m=3$ specific for Sr$_2$CuO$_3$ we find
(neglecting numerical coefficients of order one):
\begin{eqnarray}
\label{lam_typ1}
&&\widetilde \lambda \sim
\frac{Ja}{T_m}\left(\frac{T}{T_m}\right)^2\approx
20a
\left(\frac{T}{T_m}\right)^2
 \
\mbox{for}\ \ \ T<\widetilde\Theta_D,
\nonumber\\
&&\widetilde\lambda \sim \ a
\frac{J}{T_m}\frac{\widetilde\Theta_D}{T_m}\frac{T}{T_m}
\approx 60 a \frac{T}{T_m} \ \ \ \
\mbox{for}\ \ \ T>\widetilde\Theta_D .
\end{eqnarray}
The ``typical'' $k$-dependent mean free path is defined as:
\begin{eqnarray}
\label{l_typ1}
\widetilde\ell \equiv \ell (\widetilde k,T)\equiv
\frac{\ell_{\rm imp}(T)}{2}\approx 150 a\frac{T}{T_m}.
\end{eqnarray}
Therefore, $\widetilde\ell>\widetilde \lambda$ is always fulfilled
for the phonon-dominated regime.

{\it $T<T_m$ case.} \ \
In the impurity-dominated regime ``typical'' spin boson has an energy
$v|k|\sim T$. Thus, the ``typical'' wavelength is:
\begin{eqnarray}
\label{lam_typ2}
\widetilde\lambda \simeq \ a
\frac{J}{T_m}\frac{T_m}{T}\approx 20 a \frac{T_m}{T}.
\end{eqnarray}
The ``typical'' $k$-dependent mean free path is simply a cut-off length:
\begin{eqnarray}
\label{l_typ2}
\widetilde\ell = \ell_{\rm imp}(T)\approx 300 a\frac{T}{T_m}.
\end{eqnarray}
Thus, $\widetilde\ell=\widetilde \lambda$ at $T_b\approx T_m/4$ and
the quasiclassical approximation is valid above this temperature.

\section{Conclusions}
\label{Conclusions}

{\it Approximations.} \ \
We would like to  review briefly the assumptions which have
allowed us to solve the problem of spin-boson thermal transport
in a reasonably simple analytical form.

The main assumption is the smallness of the phonon relaxation
time as compared to the spin-phonon scattering time. This is justified
since the spin-lattice coupling is generically smaller than the phonon-phonon
coupling. This also allows us to restrict ourselves with the lowest
order in the spin-phonon coupling within the Boltzmann equation formalism.
In addition, in our case $c \ll v$. Thus, at a given temperature
the phonon
population is much higher than that of the spin boson.
This also implies higher relaxation rate for the phonons.
If the above is true then the momentum relaxation of spin bosons can be
viewed as a two-stage process with a bottle-neck. First, the momentum waits
the longest time for a spin-boson$-$phonon collision to get transfered
to the phonon subsystem. Second, the phonons quickly dissipate the
transferred momentum. The first stage is the bottle-neck. Therefore, it
controls the rate of the spin boson momentum dissipation. This is why we
do not need to know microscopic details about the phonon-phonon collisions.
It is enough to know that these collisions occur more often
than the spin-boson scattering on the phonons.

The factor which might lead to violation of our assumption is exponential
weakness of the phonon Umklapp relaxation at low enough temperatures.
The Umklapp relaxation rate
contains a characteristic exponent $e^{-b\Theta_D/T}$, $b$ is a constant of
order of unity.\cite{Ziman} Therefore, our approximation breaks down at very
low temperature. Roughly speaking, this temperature corresponds to the
point where the smallness of the spin-lattice coupling is comparable to
the smallness of $e^{-b\Theta_D/T}$. In
such a situation the bottle-neck scheme described above will cease to work
and it will be necessary to include a detailed description of the
phonon-phonon collisions as well.
In practice, however, the impurity scattering will become relevant at
much higher temperatures. Our theory shows a very weak impurity
concentration dependence of the temperature below which impurities
control the transport: $T_m\propto n^{1/6}$. This implies that even a
very weak disorder will result in a substantial $T_m$.

We have verified our assumptions using the phenomenological
parameters from the spin-chain material Sr$_2$CuO$_3$ and obtained
that the two-stage process approximation is very well justified, see
Sec. \ref{kappa}.

 We would like to note that the problem of the spin
transport in spin chains coupled to 3D phonons has been studied
within a different approach\cite{Andrei,Andrei1} and gives a
different temperature dependence for the spin thermal
conductivity: $\kappa_s\propto \exp(T^*/T)$. The essential
difference between our approaches is our treatment of the
phonon-bath as a source of the momentum dissipation. If this
process is neglected the ``Normal'' transfer of the momentum from
spins to phonons does not lead to degrading of the thermal
current. Then, it is only the ``direct'' spin-phonon Umklapp
processes which will be contributing to the transport relaxation
rate with a characteristic exponential dependence on temperature.
However, this could only work if the phonon system would be very
reluctant to dissipate the momentum, which can only happen at the
temperatures of the order of the ``phonon peak'' (~30K for
Sr$_2$CuO$_3$), well below the temperature of the maximum in the
spin part of $\kappa$ ($T_m=80$K).

{\it Final remarks.} \ \
In this paper we have studied the problem of anomalous heat transport
in quasi-1D spin-chain systems coupled to the 3D phonon
environment in the presence of weak disorder. We have
derived a microscopic model
of 1D bosonic spin excitations interacting with phonons
and impurities using bosonization approach. We have considered the spin-phonon
scattering within the Boltzmann equation for the
spin-boson distribution function. This equation has been solved
in the limit of weak spin-lattice coupling and fast spin-boson
excitations. Assuming that the phonon relaxation time $\tau_{pp}$
due to phonon-phonon scattering  is much
shorter than the spin-phonon relaxation time $\tau_{sp}$ we have obtained
the transport relaxation rate for the spin bosons
$\tau_{tr}^{-1}\approx\tau_{sp}^{-1}\propto k^2 T^3$ for  $T\ll
\Theta_D$ and $\propto k^2 T$ for $T\agt\Theta_D$.
We have shown that within our model the spin-phonon scattering alone
is insufficient to render the thermal conductivity finite because
the spin-phonon relaxation mechanism  becomes
ineffective for the spin bosons at low energies.
On the other hand, the 1D impurity scattering provides a natural cut-off
scale for the low-energy spin bosons and results in the
momentum-independent scattering rate $\tau_{imp}^{-1}\propto
T^{-1}$.
Thus, the impurity scattering removes the divergence of the conductivity.
We have calculated the thermal conductivity as a function of
temperature and have shown that
the low temperature transport is dominated by the impurity scattering while
the high temperature transport is limited by both the  impurity
scattering and the spin-phonon collisions.
Our main results are in a very good quantitative agreement with the
available experimental data.

In implementing our approach we have also obtained an insight into
various microscopic details of the problem.
This has allowed us to formulate several predictions and
suggest future experiments as well, see Sec. \ref{kappa}.
One of the predictions is the saturation of the spin-boson
thermal conductivity at high temperatures:
a non-trivial result due to impurity and spin-phonon
scatterings and 1D nature of spin-bosons.

Further studies in the thermal conductivity of other spin systems
using our approach are anticipated.


\begin{acknowledgments}
We would like to thank N. Andrei for useful conversations, E.
Orignac for communications, and C. Yu for insightful comments
which lead us to writing Sec. \ref{meanfp}. We are also grateful
to A. Sologubenko and C. Hess for sending experimental data. This
work was supported by DOE under grant DE-FG02-04ER46174, and by
the ACS Petroleum Research Fund. A.V.R. is grateful to the Dynasty
Foundation of Dmitrii Zimin for support of this research.

\end{acknowledgments}


\appendix
\begin{widetext}
\section{Spin-phonon collision integral}
\label{app_A}

In this Appendix we will present a detailed derivation of the relaxation time
approximation for the spin boson collision integral.

The integral over
${\bf P}$ in $S^{(1)}$, Eq. (\ref{S1}), can be evaluated explicitly.
Let us denote the
integrands of (\ref{S1}) other than delta-functions by
$h(P_\|,\omega_{\bf P})$. Then, the integration over ${\bf P}$ can be
performed giving the following answer:
\begin{eqnarray}
&&\int_{\bf P} \delta(\omega_{k'} + \omega_{{\bf P}} - \omega_k)
\delta(k' + P_\| - k) h(P_\|, \omega_{\bf P})=\frac{\omega_k -
\omega_{k'}}{8\pi^3 c^2} h(k-k', \omega_k - \omega_{k'}) \theta(|k| - |k'|)
\theta(x_0) \theta(\Theta_D^2 - x_0), \label{Pint}\\
&&\mbox{with}\ \ \ \ \ \
x_0 = v^2 (|k| - |k'|)^2 - c^2(k-k')^2, \label{x}
\end{eqnarray}
where $\theta(x)$ is a  step function and  $\Theta_D$ is the
Debye temperature.

It is possible to act in the similar fashion for $S^{(2)}$ as well. The
result of ${\bf P}$ integration for this part of the collision integral
differs by permutation of $k$ and $k'$.

The step-functions
$\theta$ in (\ref{Pint}) impose restrictions on the possible scattering
states of the spin
boson with the given momentum $k$. These restrictions are consequences of
the conservation laws. Because of this step-functions the integration
variable $k'$ in Eqs. (\ref{S1}) and (\ref{S2}) for the collision
integrals $S_{k\ell}^{(1,2)}$ is bound to an interval
${\cal C}_\ell^{(1,2)}$ smaller than the
Brillouin zone. This interval, in general, depends on $k$. To find ${\cal
C}_\ell^{(1)}$ we must solve:
\begin{eqnarray}
0 < x_0 < \Theta^2_{D\ell},
\end{eqnarray}
for $|k| > |k'|$.
Graphical solution of these inequalities for $k>0$ is presented in
Fig. \ref{fig_app} as a lightly shaded area on $(k,k')$ plane. In this Figure
lines $OA$ and $OB$ are bisectors $k=\pm k'$. Although not shown in
Fig. \ref{fig_app}, $k<0$ portion of the graphical solution can be obtained
by an inversion of Fig. \ref{fig_app} with respect to the point $O$.

Equivalently, we can express the solution as the
following choice of the integration interval ${\cal C}_\ell^{(1)}$ for $|k|
> \Theta_D/v$:
\begin{eqnarray}
{\cal C}_\ell^{(1)} = \cases{
[k, k-\Theta_{D\ell}/v] \cup [-k+\sqrt{4(c_\ell/v)^2 k^2 +
(\Theta_{D\ell}/v)^2}, -(1-2c_\ell/v)k]& for $k > 0$,\cr
[ -(1-2c_\ell/v)k, -k-\sqrt{4(c_\ell/v)^2 k^2 +(\Theta_{D\ell}/v)^2}]
\cup [k+\Theta_{D\ell}/v, k] & for $k < 0$,
}
\end{eqnarray}
where we use $[k_{min},k_{max}]$ to denote the interval from $k_{min}$ to
$k_{max}$.  For $|k| < \Theta_{D\ell}/v$:
\begin{eqnarray}
{\cal C}_\ell^{(1)} = \cases{ [k, -(1-2c_\ell/v)k]& for $k > 0 $,\cr
[-(1-2c_\ell/v)k, k]& for $k < 0$.  }
\end{eqnarray}
For the contour  ${\cal C}^{(2)}_\ell$ in the integral for
$S_{k\ell}^{(2)}$:
\begin{eqnarray}
{\cal C}_\ell^{(2)} = \cases{
[k+\Theta_{D\ell}/v, k] \cup [-(1+2c_\ell/v)k, -k - \Theta_{D\ell}/v]& for
$k > 0$,\cr
[k - \sqrt{4(c_\ell/v)^2 k^2 + (\Theta_{D\ell}/v)^2}, (1+2c_\ell/v)k] \cup
[k, k - \Theta_{D\ell}/v]& for $k < 0$.}
\end{eqnarray}
The $k>0$ part of this solution corresponds to the darker shaded area in
Fig. \ref{fig_app}.

\begin{figure}[t]
\includegraphics[width=6cm]{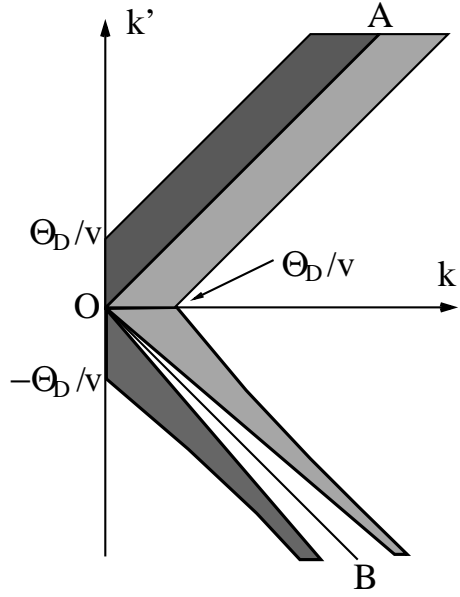}
\caption{This diagram shows the regions in the momentum space, allowed
  by the momentum and energy conservation, for the spin boson
  scattered from the state $k$ to the state $k'$.
The processes in which the original spin
boson decays into another spin boson and emits the phonon are
  represented by the
light gray area. For them $|k'| < |k|$ as dictated by the
energy conservation.
The dark gray area corresponds to the processes of phonon
absorption by the spin boson. For them $|k'| > |k|$. Lines $OA$ and $OB$ are
bisectors $k' =\pm k$.}
\label{fig_app}
\end{figure}

Keeping in mind the above results
we rewrite the collision integral in the following form:
\begin{eqnarray}
S_{k} &=& \sum_\ell S_{k\ell},\label{S}\\
S_{k\ell}&=&{\cal A}_\ell \int_{{\cal C}_\ell} \frac{dk'}{2\pi}
\left|\frac{kk' (k-k')^2 }{e^{(|k| - |k'|)/k_T}-1}\right|
({\bm \xi}_{k\ell})_x^2
\left\{ \frac{e^{|k|/k_T} - 1}{e^{|k'|/k_T} - 1} f_{k}^1
- \frac{e^{-|k'|/k_T} - 1}{e^{-|k|/k_T} - 1} f_{k'}^1 \right\},
\label{Sell}
\end{eqnarray}
where
\begin{eqnarray}
&&{\cal A}_\ell=\frac{V_0}{64\pi^3 m_i} \left(\frac{g_{sp}}{c_\ell}
\right)^2,\\
\mbox{and} &&k_T=T/v \ .
\end{eqnarray}
The contour ${\cal C}_\ell$ is the union ${\cal C}_\ell^{(1)} \cup
{\cal C}_\ell^{(2)}$. For $|k| < \Theta_{D\ell}/v$ we must integrate
between $\pm(|k| + \Theta_{D\ell}/v)$. For $|k| > \Theta_{D\ell}/v$ we
must integrate between $k \pm \Theta_{D\ell}/v$ and between $-k \pm
\sqrt{4(c_\ell/v)^2 k^2 + (\Theta_{D\ell}/v)^2}$. In both cases a window
of the width $4c_\ell |k|/v$ centered around $k' = -k$ must be deleted
from the integration interval. In Fig. \ref{fig_app}
one can identify this window with the
unshaded area around line $OB$.

We start by making the following observation. In general, the intervals
${\cal C}_\ell$ in (\ref{Sell}) differ for different
polarizations $\ell=l,t$: the size of the gap around $k' = -k$ depends
on the value of $c_\ell$, so do $\Theta_{D\ell}/v$ which specify the outer
limits of the integration intervals. However, we will show later that these
variations in ${\cal C}_\ell$ do not change the integrals (\ref{Sell})
significantly.
Therefore, it is convenient to introduce ${\cal C}_0 = {\cal C}_l \cap
{\cal C}_t$ and $\delta{\cal C}_\ell = {\cal C}_\ell - {\cal C}_0$. Using
this notation we write:
\begin{eqnarray}
S_\ell = \int_{{\cal C}_\ell} \ldots = \int_{{\cal C}_0} \ldots +
\int_{\delta{\cal C}_\ell} \ldots.
\end{eqnarray}
As it was mentioned the integrals over $\delta{\cal C}_\ell$ in the
above expressions are small. Hence, it is permissible to integrate over
${\cal C}_0$ in both $S_t$ and $S_l$. In such a situation the summation over
polarization $\ell$ can be done easily since in both $S_t$ and $S_l$ only the
product ${\cal A}_\ell ( {\bm \xi}_\ell)^2_x$ depends on
polarization.
Thus:
\begin{eqnarray}
{\widetilde{\cal A}}
= \sum_\ell {\cal A}_\ell \left( {\bm \xi}_\ell\right)^2_x =
{\cal A}_t,
\end{eqnarray}
where ${\bm \xi}_{lx} = c_l |k - k'|/v ||k| - |k'||$ and
${\bm \xi}_{tx} = \sqrt{ 1 - (c_t |k - k'|/v ||k| - |k'||)^2}$.
Therefore, the collision integral itself is equal to:
\begin{eqnarray}
S_{k}&=&{\widetilde{\cal A}} \int_{{\cal C}_0} \frac{dk'}{2\pi}
\left|\frac{kk' (k-k')^2 }{e^{(|k| - |k'|)/k_T}-1}\right|
\left\{ \frac{e^{|k|/k_T} - 1}{e^{|k'|/k_T} - 1} f_{k}^1
- \frac{e^{-|k'|/k_T} - 1}{e^{-|k|/k_T} - 1} f_{k'}^1 \right\}.\label{Sapp}
\end{eqnarray}
The first term of this integral can be calculated for small $|k|$:
\begin{eqnarray}
{\widetilde{\cal A}} f_k^1 \int_{{\cal C}_0} \frac{dk'}{2\pi}
\left|\frac{kk' (k-k')^2}
{e^{(|k| - |k'|)/k_T} - 1} \right| \frac{e^{|k|/k_T} - 1}{e^{|k'|/k_T} -1}
\approx{\widetilde{\cal A}} f_k^1 \frac{k^2}{k_T}
\int_{-\Theta_D/v}^{\Theta_D/v}
\frac{dk'}{2\pi} \frac{|k'|^3}{2\cosh(k'/k_T)-2}
= \frac{f^1_k }{\tau_{\rm sp} (k)},\label{tau}
\end{eqnarray}
which gives the spin-boson transport relaxation time:
\begin{eqnarray}
\tau_{\rm sp} = \cases{{v^3}/{{\cal A} T^3 k^2}& for $T \ll \widetilde
\Theta_D$\cr
{v^3}/{{\cal A} \widetilde \Theta^2_D T k^2}& for $T \gg \widetilde
\Theta_D$\cr}
\end{eqnarray}
\begin{eqnarray}
\mbox{with} \ \
{\cal A}=\frac{I_1(\infty)}{2\pi}{\widetilde{\cal A}} \approx
2.3 \widetilde{\cal A}, \quad \widetilde \Theta_D =
\frac{\Theta_D}{\sqrt{I_1(\infty)}} \approx 0.25 \Theta_D,
\label{tau_tr}
\end{eqnarray}
where $I_1(z)$ is defined in Eq. (\ref{gamma0}). One can see from
Eq. (\ref{tau}) that the dominant scattering process is the absorption
of a phonon with the characteristic energy $\sim T$ as this energy
range provides the major contribution to the integral.

If we neglect the second term of (\ref{Sapp}) as well as the corrections to
the collision integral coming from $\delta{\cal C}_\ell$ the function $f_k^1$
can be found right away. For example, if $T \ll \widetilde \Theta_D$:
\begin{eqnarray}
f_k^1 =
\left\{ \frac{v^4}{{\cal A} T^4 |k|}
\frac{\partial}{\partial k} f^0_k (T) \right\} \partial_x T.\label{f1}
\end{eqnarray}
This is the expression for odd (in $k$) part of the spin boson distribution
function. Using it we estimate the omitted term of (\ref{Sapp}) and show that
this term is small:
\begin{eqnarray}
-{\widetilde{\cal A}}
 \int_{{\cal C}_0} \frac{dk'}{2\pi} \left|\frac{kk' (k-k')^2}
{e^{(|k| - |k'|)/k_T} - 1} \right|
\frac{e^{-|k'|/k_T} - 1}{e^{-|k|/k_T} -1}f_{k'}^1 =
-{\widetilde{\cal A}} \int_{{\cal C}_0} \frac{dk'}{2\pi} \left|\frac{kk'}
{e^{(|k| - |k'|)/k_T} - 1} \right| \left[ k^2 - 2kk' + k'^2 \right]
\frac{e^{-|k'|/k_T} - 1}{e^{-|k|/k_T} -1}f_{k'}^1.\nonumber
\end{eqnarray}
Of three terms in the square brackets only the second gives non-vanishing
contribution to the integral. Two others do not contribute because they are
even in $k'$ and function $f^1_{k'}$ is odd. Therefore:
\begin{eqnarray}
-{\widetilde{\cal A}} \int_{{\cal C}_0}
\frac{dk'}{2\pi} \left|\frac{kk' (k-k')^2}
{e^{(|k| - |k'|)/k_T} - 1} \right|
\frac{e^{-|k'|/k_T} - 1}{e^{-|k|/k_T} -1}f_{k'}^1
&\approx& \frac{2k}{\alpha k_T^3} \partial_x T
\int_{-\Theta/v}^{\Theta/v}
\frac{dk'}{2\pi} k' \frac{\partial}{\partial k'} f^0_{k'}
\approx
\left\{ \frac{2k}{\pi \alpha k_T^2} \log \frac{k_{min}}{k_c} \right\}
\partial_x T, \label{ax}
\end{eqnarray}
where $k_{min} = \min\{k_T; \Theta_D/v\}$.
Since the integral in (\ref{ax}) diverges at small $|k'|$ we must introduce
small cut-off parameter $k_c$. Such a cut-off
is associated with the impurity scattering.
It is equal to the momentum at which the impurity scattering rate becomes
comparable with $\tau_{\rm sp}^{-1}$. For now it is enough to assume that
$k_c$ is not too small so we can neglect the logarithm in the above formula.
The expression (\ref{ax}) is to be compared with (\ref{tau}). For
$T \ll \widetilde \Theta_D$:
\begin{eqnarray}
\frac{f_k^1}{\tau_{sp}} \approx
\frac{1}{k}\partial_x T \gg
\left\{ \frac{2|k|}{\pi \alpha k_T^2} \log \frac{k_T}{k_c} \right\}
\partial_x T \Leftrightarrow |k|^2 \ll k_T^2.
\end{eqnarray}
In other words, at low temperatures the second term of (\ref{Sapp}) is small
as long as $|k| \ll k_T$. The same is true for larger temperature
($T \agt \widetilde \Theta_D$) as well.

We also need to estimate the integrals over $\delta{\cal C}_\ell$ to
demonstrate that they are small. First,
let us consider the integral over $\delta{\cal C}_t$. This interval is
localized near $k' = -k$:
\begin{eqnarray}
\delta{\cal C}_t = [-(1 + 2c_l/v)k, -(1 + 2c_t/v)k] \cup
 [-(1 - 2c_t/v)k, -(1 - 2c_l/v)k].
\end{eqnarray}
It is non-vanishing as long as $c_l > c_t$. It is necessary to perform the
integral (\ref{Sell}) with $\ell = t$ over $\delta{\cal C}_t$:
\begin{eqnarray}
&&{\cal A}_t \int_{\delta{\cal C}_t} \frac{dk'}{2\pi}
\left|\frac{kk' (k-k')^2 }{e^{(|k| - |k'|)/k_T}-1}\right|
({\bm \xi}_{kt})_x^2
\left\{ \frac{e^{|k|/k_T} - 1}{e^{|k'|/k_T} - 1} f_{k}^1
- \frac{e^{-|k'|/k_T} - 1}{e^{-|k|/k_T} - 1} f_{k'}^1 \right\} \\
&&\approx
4k^4 {\cal A}_t \left\{ f_k^1 - f_{-k}^1 \right\} \int_{\delta{\cal C}_t}
\frac{dk'}{2\pi} \left( 1 - \frac{c_t^2 (k - k')^2}{v^2 (|k| - |k'| )^2}
\right) \left| \frac{1}{e^{(|k| - |k'|)/k_T} - 1} \right| \nonumber
\\
&&\approx \frac{2{\cal A}_t}{\pi}
T k^4 \left\{ f_k^1 - f_{-k}^1 \right\} \int_{\delta{\cal C}_t}
dk' \frac{v^2(|k| - |k'|)^2 - c_t^2 (k - k')^2}
{v^3 ||k| - |k'| |^3} . \nonumber
\end{eqnarray}
The absolute value of the last expression can be bounded from above by the
following:
\begin{eqnarray}
\frac{{\cal A}_t}{4\pi c_t^3}
T |k| \left| f_k^1 - f_{-k}^1 \right| \int_{\delta{\cal C}_t}
dk' \left(v^2(|k| - |k'|)^2 - c_t^2 (k - k')^2\right) \approx
\frac{4{\cal A}_t}{\pi}
k_T k^4 \left| f_k^1 - f_{-k}^1 \right| \left(
\frac{1}{3}  \frac{(c_l - c_t)^3}{c_t^3} +
 \frac{(c_l - c_t)^2}{c_t^2} \right).
\end{eqnarray}
The latter expression can be estimated as $\propto {\cal A}_t k_T k^4
|f_k^1|$. This quantity is smaller than (\ref{tau}) for $|k| < k_T$ as long
as $(k/k_T)^2\ll 1$.

Second, we estimate the integral (\ref{Sell}) with $\ell=l$ over
$\delta{\cal C}_l$:
\begin{eqnarray}
\delta{\cal C}_l = [|k| + \Theta_{Dl}/v, |k| + \Theta_{Dt}/v] \cup
[-|k| - \Theta_{Dt}/v, -|k| - \Theta_{Dl}/v].
\end{eqnarray}
It is clear without extensive calculations that both terms of (\ref{Sell})
are proportional to $({\bm \xi}_{kl})^2_x \approx (c/v)^2 \ll 1$.
Thus, the integral over this interval does not contribute significantly to
the collision integral of the spin bosons.

Therefore, we can conclude that for small $T$ the collision integral can be
approximated by the expression (\ref{tau}).

\section{Impurity scattering}
\label{app_B}

Now we can start evaluating the self-energy corrections to the spin boson
Green's function. Our first step is the calculation of the Matsubara Green's
function ${\cal D}(\tau)$. Since the perturbation is an exponential of
the bosonic field $\tilde \Phi$ rather than a polynomial we present our
calculation in details. The second order correction to the Green's function
is given by:
\begin{eqnarray}
{\cal D}_k(\tau) - {\cal D}_{0k}(\tau) \approx
\frac{\delta J_{\rm imp}^2}{2\pi^2 }
\int d\tau ' d\tau ''
\left< \left( b^\dagger_k(\tau) + b^{\vphantom{\dagger}}_{-k}(\tau) \right)
{e}^{-{i} \sqrt{2\pi} \tilde\Phi(x_0,\tau')}
{e}^{{i} \sqrt{2\pi} \tilde \Phi(x_0 ,\tau'')}
\left( b^{\vphantom{\dagger}}_{k'}(0) + b^\dagger_{-k'}(0) \right)\right>_{\rm
con-\atop nected}.\label{D2}
\end{eqnarray}
The triangular brackets stand for the Matsubara time-ordered
averaging, and $x_0$ is the impurity position. The
bare boson Green's function
\begin{eqnarray}
{\cal D}_{0k}(\tau) = \left< \left( b^\dagger_k (\tau) +
b^{\vphantom{\dagger}}_{-k}(\tau) \right)\left( b^{\vphantom{\dagger}}_k (0)
+ b^\dagger_{-k}(0) \right) \right>
\end{eqnarray}
is a Fourier transform of
${\cal D}_{0k,i\omega} = 2\omega_k/(\omega^2 + \omega_k^2)$. To calculate the
expression in the triangular brackets in Eq. (\ref{D2})
we must expand both exponents into
Taylor series
\begin{eqnarray}
\sum_{n,m} \frac{(-i\sqrt{2\pi})^n(i\sqrt{2\pi})^m}{n! m!}
\left< \left( b^\dagger_k(\tau) + b^{\vphantom{\dagger}}_{-k}(\tau) \right)
\tilde\Phi^n (x_0,\tau') \tilde \Phi^m (x_0 ,\tau'')
\left( b^{\vphantom{\dagger}}_{k'}(0) + b^\dagger_{-k'}(0) \right)\right>_{\rm
con-\atop nected}
\end{eqnarray}
and apply Wick's theorem. The external operators $b,\ b^\dagger$ can be
contracted with fields $\tilde \Phi$ in four possible ways schematically
shown in Fig. \ref{imp_scatt}:
(a) both external operators contracted with the
monomial $\tilde \Phi^n (x_0, \tau')$, (b) both external operators
contracted with the monomial $\tilde \Phi^m (x_0, \tau'')$, and (c) and
(d) one external operator contracts with
$\tilde \Phi^n (x_0, \tau')$ while another contracts with
$\tilde \Phi^m (x_0, \tau'')$. Each contraction with the external operator
acts on a monomial as a derivative with
respect to the field $\tilde \Phi$ times $e^{ikx_0}{\cal D}_{0k}/\sqrt{L|k|}$
or $e^{-ik'x_0}{\cal D}_{0k'}/\sqrt{L|k'|}$. Effectively, contraction with
an external operator is equivalent to differentiation with respect to
$\tilde \Phi$ times some $c$-number. Because of that one can derive:
\begin{eqnarray}
-\frac{\pi{e}^{{i} ( k - k') x_0 }}
{L\sqrt{|kk'|}} \bigg\{{\cal D}_{0k}(\tau') {\cal D}_{0k'} (\tau - \tau')
+ {\cal D}_{0k}(\tau'') {\cal D}_{0k'} (\tau - \tau'')
-{\cal D}_{0k} (\tau') {\cal D}_{0k'}(\tau - \tau'') -
{\cal D}_{0k} (\tau''){\cal D}_{0k'} (\tau - \tau')
\bigg\} \label{aux}\\
\times \left< {e}^{-{i} \sqrt{2\pi} \tilde\Phi(x_0,\tau')}
{e}^{{i} \sqrt{2\pi} \tilde \Phi(x_0 ,\tau'')} \right>. \nonumber
\end{eqnarray}
The first term in the curly brackets corresponds to the diagram (a) in
Fig. \ref{imp_scatt}, the second corresponds to the diagram (b) and so on.
The gray bubbles in Fig. \ref{imp_scatt} correspond to
$\langle e^ {-i \sqrt{2\pi} \tilde \Phi (0, \tau') }
e^ {i \sqrt{2\pi} \tilde \Phi (0, \tau'')} \rangle$.

As we see from (\ref{aux}), the impurity scattering explicitly
violates the momentum conservation. It will be restored once we average
over the impurity positions. Such impurity averaging leads to a
simple change in the above expression:
$\langle\langle e^{i(k-k')x_0}...\rangle\rangle_{imp}/L=n
\delta_{k,k'}/a$, where $n$ is
the dimensionless
 impurity concentration. After that the impurity position $x_0$ is
arbitrary and can be put to zero. This yields:
\begin{eqnarray}
&&{\cal D}_k (\tau) - {\cal D}_{0k}(\tau) \approx \frac{1}{2}
\int d\tau' d\tau'' \bigg\{ {\cal D}_{0k}(\tau - \tau')
{\cal D}_{0k} (\tau'') + {\cal D}_{0k} (\tau - \tau'') {\cal D}_{0k}
(\tau')
\nonumber\\
&&\phantom{{\cal D}_k (\tau) - {\cal D}_{0k}(\tau) \approx}
- 2 {\cal D}_{0k} (\tau - \tau'') {\cal D}_{0k} (\tau'')
\bigg\} \Sigma_k (\tau' - \tau''),\\
&&\Sigma_k (\tau' - \tau'')= \frac{\delta J_{\rm imp}^2 n}{a\pi  |k|}
\left<{e}^{-{i} \sqrt{2\pi} \tilde\Phi(0,\tau')}
{e}^{{i} \sqrt{2\pi} \tilde \Phi(0 ,\tau'')} \right>.
\end{eqnarray}
With the help of the standard formula for free bosonic field:
\begin{eqnarray}
\langle e^{ \alpha \tilde \Phi(x,\tau) }e^{ - \alpha \tilde \Phi(x',\tau') }
\rangle = e^{{\alpha^2} \langle (\tilde \Phi(x,\tau) -
\tilde \Phi(x',\tau'))^2 \rangle /2}
\end{eqnarray}
the above equation can be rewritten in the $\omega$-space as:
\begin{eqnarray}
&&{\cal D}_{k,i\omega} - {\cal D}_{0k,i\omega} \approx
{\cal D}_{0k,i\omega}^2 \Sigma_{k,i\omega},\label{DS}\\
&&\Sigma_{k,i\omega} = \frac{\delta J_{\rm imp}^2 n}{a\pi  |k|}
\int d\tau \left({e}^{i\omega \tau} - 1 \right)
{e}^{ -\pi \left< ( \tilde \Phi (0,\tau) -
\tilde \Phi (0,0) )^2 \right> }. \label{Sigma}
\end{eqnarray}
The average of the bosonic fields can be evaluated as follows
\begin{eqnarray}
g(\tau) &=& - \left< \left(\tilde \Phi (0,\tau) - \tilde \Phi (0,0)
\right)^2 \right> = 2 \left< \tilde \Phi(0,\tau) \tilde \Phi(0,0) \right>
- 2\left< \tilde \Phi(0,0)^2 \right>\\
&=& 2 v T\sum_{\omega\ne 0} \int_k \frac{ {e}^{{i} \omega \tau} - 1}
{\omega^2 + v^2 k^2} {e}^{-|k|a} =
2 T \sum_{\omega\ne 0} \left( {e}^{{i} \omega \tau} - 1 \right)
I(\omega),\nonumber\\
\mbox{with}\ \
I(\omega) &=& \frac{1}{\pi |\omega|} \int_0^{+\infty} \frac{{\rm
e}^{-a|\omega|x/v}}{1 + x^2} dx.
\end{eqnarray}
It is possible to evaluate the integral $I(\omega)$ for small ($|\omega|
\ll J \sim v/a$) and large ($|\omega| \gg J \sim v/a$) Matsubara frequency
$\omega$:
\begin{eqnarray}
I(\omega) = \cases{ {1}/{2|\omega|} & for $|\omega| \ll J$ \cr
{v}/{\pi \omega^2 a} & for $|\omega| \gg J$.\cr}
\end{eqnarray}
The fact that at large $|\omega|$ function $I(\omega)$ decays as an
inverse $\omega^2$ guarantees the convergence of the Matsubara sum.
Unfortunately, it is impossible to continue our derivation analytically due
to complex properties of $I$. To overcome this difficulty we utilize a
trick. Rather than using the actual form of  $I(\omega)$ we put a
different function in its place:
\begin{eqnarray}
\tilde I(\omega) = \frac{e^{-|\omega|/J}}{2|\omega|} \ .
\end{eqnarray}
 Such a replacement preserves $|\omega| \ll J$ properties of the
sum but radically improves convergence at large $\omega$.
Since we are not interested in the high energy ($\sim J$), short time
($\sim 1/J$) properties of $\Sigma$ this procedure is justified.
Now the Matsubara summation can be done and it leads to:
\begin{eqnarray}
g(\tau) = T \sum_{\omega\ne 0} \frac{ e^{i\omega\tau - |\omega|/J} -
e^{-|\omega|/J}} {|\omega|} = \frac{1}{2\pi}
\log \frac{ \left( 1 - e^{-2\pi T/J} \right)^2}
{\left( 1 + e^{-4\pi T/J} - 2 e^{-2\pi T/J} \cos 2\pi T\tau \right)}.
\end{eqnarray}
In deriving this formula we used the identity:
\begin{eqnarray}
\sum_{\omega > 0} \frac{e^{\gamma \omega}}{\omega} =
- \frac{1}{2\pi T} \log \left( 1 - e^{2\pi T \gamma} \right),
\end{eqnarray}
which is correct for any complex $\gamma$, ${\rm Re}\ \gamma < 0$. To prove
this formula one can expand the logarithm in
Taylor series with respect to powers of
$e^{2\pi T \gamma}$.

Once $g(\tau)$ is calculated it must be substituted into (\ref{Sigma}). Now
we have to evaluate the Fourier integral:
\begin{eqnarray}
F(\omega) = \int_0^\beta d\tau
\frac{ \left( e^{i\omega \tau} - 1 \right) \left( 1 - e^{-2\pi T/J} \right)}
{\sqrt{ 1 + e^{-4\pi T/J} - 2 e^{-2\pi T/J} \cos 2\pi T\tau }}.
\end{eqnarray}
This integral can be transformed into an integral over a unit circle in the
complex plane:
\begin{eqnarray}
F = \frac{\sqrt{s}}{2\pi i T} \oint  dz
\frac{z^m-1}{\sqrt{z((2+s)z - z^2 -1)}}, \label{F}
\\
s = 4\pi^2 T^2/J^2 \ll 1, \quad z = e^{2\pi i T \tau}, \quad
m = \frac{\omega}{2\pi T} - {\rm integer}.
\end{eqnarray}
The polynomial under the square root has three zeros: $z_0 = 0$ and
$z_{1,2} \approx 1 \pm \sqrt{s}$. Our branch-cut consists of two components:
the first component stretches from $0$ to $1-\sqrt{s}$, the second component
stretches from $1+\sqrt{s}$ to $+\infty$. Such a choice of the branch-cut
ensures that the integration path, the unit circle, does not cross the
branch-cut, see Fig. \ref{contour}.
\begin{figure}[t]
\includegraphics[width=6cm]{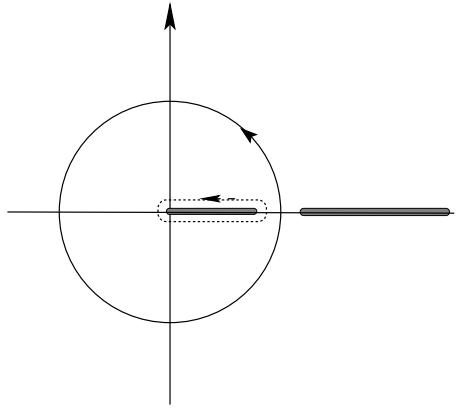}
\caption{The transformation of the  integration path in the complex
  plane for Eq. (\ref{F}) is shown. The thick gray lines represent
the branch cuts. The unit circle is
transformed into a new contour circumventing one of the branch
cuts. This new contour is shown by the dash line.}
\label{contour}
\end{figure}
It is possible now to change the integration path. The new
integration path runs counterclockwise along the sides of $(0,1-\sqrt{s})$
component of the branch-cut. In turn, this integral can be transformed into an
integral over the real axis:
\begin{eqnarray}
F=\frac{\sqrt{s}}{\pi T}
\int_0^{1-\sqrt{s}} dx \frac{(x^{m}-1)\sqrt{x}} {\sqrt{ (x-1)^2 - sx }}.
\end{eqnarray}
Finally, the Matsubara self-energy equals to
\begin{eqnarray}
\Sigma_{k,i\omega} \approx \frac{\delta J_{\rm imp}^2 n}{a\pi^2 T |k|}
\sqrt{s}\int_0^{1} dx \frac{(x^m - 1)\sqrt{x}} {x-1}.
\end{eqnarray}
To obtain the retarded self-energy $\Sigma^R$ the analytical continuation
$m \rightarrow -i\omega/2\pi T$ must be done. For $\omega \ll T$ it gives:
\begin{eqnarray}
\Sigma^R_{k,\omega} \approx
-i\frac{\delta J_{\rm imp}^2 n}{2 a J |k| T} \omega.
\end{eqnarray}
Once the self-energy is calculated one can find the retarded
Green's function:
\begin{eqnarray}
D_{k,\omega} = \frac{D_{0k,\omega}}{ 1 - D_{0k,\omega} \Sigma_{k,\omega}^R}
= \frac{2\omega_k}{-\omega^2 + \omega_k^2 - 2\omega_k
  \Sigma_{k,\omega}^R} \,
\end{eqnarray}
and extract the lifetime from the position of the Green's function pole:
\begin{eqnarray}
\omega^2 = \omega_k^2 - 2\omega_k \Sigma_{k,\omega}^R \ .
\end{eqnarray}
As a result the lifetime $\Gamma$ is  given by:
\begin{eqnarray}
\Gamma = \frac{\Delta^2}{T},
\end{eqnarray}
with the auxiliary parameter $\Delta$ is defined as
\begin{eqnarray}
\Delta^2 \propto n \delta J_{\rm imp}^2
\end{eqnarray}
where $n$ is the dimensionless impurity concentration.
\end{widetext}



\begin{thebibliography}{99}
\bibitem{comment} Further clarification of this issue is
planned for the future publication.
\end{thebibliography}
\end{document}